\documentstyle[12pt]{article}
\begin{document}
\thispagestyle{empty}
\begin{flushright}
CUPP-99/1\\
\texttt{hep-ph/9901401} \\
January 1999\\
\end{flushright}
\vskip 5pt

\begin{center}
{\large {\bf NEW VARIABLES FOR NEUTRINO OSCILLATION DIAGNOSTICS
AT SUPERKAMIOKANDE AND THE SUDBURY NEUTRINO OBSERVATORY}}

\vskip 10pt

{\sf Debasish Majumdar
\footnote{E-mail address: debasish@tnpdec.saha.ernet.in}}  
and 
{\sf Amitava Raychaudhuri
\footnote{E-mail address: amitava@cubmb.ernet.in}}  

\vskip 8pt

{\em Department of Physics, University of Calcutta,\\ 
92 Acharya Prafulla Chandra Road, Calcutta 700 009, India }

\vskip 8pt

{\bf ABSTRACT}

\end{center}

The SuperKamiokande collaboration has presented results on the
observation of solar neutrinos. The Sudbury Neutrino Observatory
(SNO) is also expected to go on-line in the near future. We
propose several new variables, insensitive to the absolute flux
of the initial solar neutrino beam, which probe the shape of the
observed spectrum at these experiments and can sensitively signal
neutrino oscillations. One class of such variables involves
normalised moments of the distributions recorded at the two
facilities while another variable, specific to SNO, depends on
the integrated charged and neutral current signals.  The utility
of these variables in the context of supernova neutrinos, both
from the collapse epoch and the post-bounce era, is also
emphasised. It is shown that, notwithstanding the imprecise nature
of the information about the initial neutrino spectra from a
supernova, oscillations can be detected using these variables and
it will be possible to distinguish between the alternatives of
oscillation to a sequential neutrino {\em vis-a-vis} that to a
sterile neutrino.

\begin{center} 
\end{center}

\newpage

\section{Introduction}

Neutrino physics has moved centrestage after the recent
SuperKamiokande (SK) evidence in support of a non-zero neutrino
mass and oscillation in their atmospheric $\nu$ data \cite {ska}.
A massive neutrino has vast implications in astrophysics and
cosmology and also signals physics outside the Standard Model
(SM) \cite {numass}. Further indications of neutrino mass and the
determination of the complete mass spectrum are therefore awaited
with  much interest.

Historically, the long-standing solar neutrino problem \cite {solar}
had offered the first glimpse of the oscillation of a $\nu_e$ to
a different neutrino. This oscillation partner state could be one
of the sequential neutrinos, $\nu_\mu$ or $\nu_{\tau}$. It could
just as well be a sterile neutrino, $\nu_s$, which has no weak
interactions. If produced by neutrino oscillations, the latter will
completely escape the detectors. The inclusion of a fourth
neutrino -- sterile, in view of the LEP and SLC results -- is
suggested from the several evidences indicative of neutrino
oscillations, namely, the solar neutrino puzzle, the atmospheric
neutrino anomaly and the results of the LSND experiment
\cite{lsnd}, all of which cannot be accommodated together in a
three neutrino framework \cite{goswami}.

New, high statistics results on solar neutrinos are just becoming
available.   The huge fiducial volume of the SuperKamiokande
detector \cite {suk} has already enabled the accumulation of data
at unprecedented rates \cite{sks}, improving, and to a large extent,
corroborating  the earlier results on solar neutrinos.
Another detector of comparable size, the Sudbury Neutrino
Observatory (SNO) \cite {sno} is shortly expected to be
operational.  The latter experiment, because of its capability to
detect neutrinos {\em via} both charged current (CC) as well as
neutral current (NC) detection channels, will shed light on the
nature of the other neutrino -- sequential or sterile -- involved
in the oscillation explanation of the reduction of the solar
neutrino flux from its ``standard" value.  The large data sample
from these two experiments will provide an opportunity to examine
signals for neutrino oscillation in novel manners.

An uncertainty in drawing conclusions from the solar neutrino
data creeps in through the imperfect knowledge of the initial
neutrino flux.  Though the shape of the neutrino spectra from the
different reactions occuring in the sun (the $p$-$p$ chain, CNO
cycle, {\em etc.}) are known precisely from weak interactions and
nuclear physics, their relative and absolute normalisations
depend on the physics and astrophysics within the sun and vary
from one solar model to another \cite {bah,turck}. The Boron
neutrinos -- only which are seen at SK and SNO -- are
particularly sensitive with the absolute normalisation varying,
for example, as $T_c^{18}$, where $T_c$ is the solar core
temperature. It is therefore of interest to formulate methods to
decipher signatures of oscillation in the observed data which are
independent of this imperfect knowledge of the normalisation of
the initial flux.

In this work we examine in detail several variables which depend
on (a) the normalised moments of distributions seen at SK and SNO
and (b) on the ratios of the charged and neutral current signals
at SNO \cite{mr}.  Though these variables are sensitive to the
precisely understood shape of the Boron-neutrino spectrum, they
are independent of its absolute normalisation.  The high
statistics data from the new detectors make such a study
feasible. We illustrate how the magnitudes of the above-mentioned
variables (and their ratios) extracted from the data yield direct
information on the neutrino mixing angles and mass splittings.
We further show how the variables can be used to distinguish
whether the $\nu_e$ oscillates to a sequential neutrino or to a
sterile one.  Some similar ideas have also been advocated in ref.
\cite{blsno,prev} where the focus has been on the energy spectrum
of the scattered electron neutrino at SNO, the MSW mechanism etc.
In this work, we restrict ourselves to vacuum oscillations. At
the present time the existing evidences for solar netrino
oscillations cannot clearly distinguish between the vacuum
oscillation and MSW resonant flavour conversion alternatives. In
the latter case, the expressions for neutrino oscillation
probabilities that we use in this work will have to be
appropriately modified \cite{mr2}.

The proposed variables can also serve as useful tools to examine
neutrino signals from a supernova. Though, in this case, the
shape of the initial neutrino flux is known less precisely,
nonetheless, we show that in the event of oscillations the
variables can assume values which are beyond the range that can
be expected from this uncertainty.  Neutrinos are emitted from
two distinct epochs of a supernova explosion. In the collapse
phase, which occurs first, only $\nu_e$ produced from electron
capture are emitted while in the post-bounce era neutrinos of all
three flavours are produced. The pure $\nu_e$ nature of the
collapse phase beam -- like the solar neutrino case -- holds some
advantages but this is partially offset by the much smaller
number compared to the neutrinos emitted in the post-bounce
period. While the detection of the latter for a supernova within
a 10 kpc distance is very likely, those from the former will be
observable provided the explosion occurs within a distance of
about 1 kpc.  Some initial results on these issues have been
reported earlier \cite{mkrrs}.

In the next section we introduce the variables which we propose
and show how they can signal solar $\nu$-oscillations to a
sequential or a sterile neutrino.  In section 3 we consider
supernova neutrinos and illustrate how the variables can be
useful in that context. We end in section 4 with some
discussions.

\section{The new variables and solar neutrinos}

In this section we introduce variables insensitive to the
absolute normalisation of the initial flux which may be used as
diagnostic tools for solar neutrino oscillations at
SuperKamiokande and SNO. We consider the effect of
$\nu$-oscillations on the solar neutrino spectrum as seen at
these detectors and elaborate on the sensitivity of the new
variables. We restrict ourselves to the two-flavour oscillation
case. Mixing of more than two kinds of neutrinos will change the
expressions for the oscillation probability in a well-known
fashion. This can be readily incorporated but will introduce more
parameters in the form of additional mixing angles and mass
splittings.

\subsection{Solar Neutrino Oscillations at SuperKamiokande and SNO}

In the two-flavour case, the probability of an electron neutrino
of energy $E$ to oscillate to a neutrino of a different
type, $\nu_x$, after the traversal of a distance $L$ is:
\begin{equation}
P_{\nu_e \rightarrow \nu_x} = \sin^2(2\vartheta) \sin^2 \left
(\frac {\pi L} {\lambda} \right )
\label{ex}
\end{equation}
where $\vartheta$ is the mixing angle. $\lambda$ is the oscillation length
given, in terms of the mass-squared difference $\Delta$, by:
\begin{equation}
\lambda = 2.47 \left (\frac
{E} {\rm MeV} \right ) \left (\frac {\rm eV^2} {\Delta}\right
)\; {\rm metre}
\end{equation}
From probability conservation: $P_{\nu_e \rightarrow \nu_e} =  1 -
P_{\nu_e \rightarrow \nu_x}$.

In the above, $\nu_x$ can be either a sequential neutrino,
$\nu_{\mu}$ or $\nu_{\tau}$, or a sterile neutrino, $\nu_s$. The
difference between these two cases will manifest itself at the
detectors as follows. At SuperKamiokande the neutrinos are
detected {\em via} $\nu - e$ scattering. For the $\nu_e$ there
are contributions through both CC and NC weak interactions. If
neutrino oscillations are operative then in case a sequential
neutrino is produced it will contribute to the signal only
through the NC interactions (roughly one sixth of the $\nu_e$
case) while a sterile neutrino will be entirely missed by the
detector. At SNO the electron neutrinos will be detected through
(a) CC as well as (b) NC interactions. If oscillations to
sequential neutrinos occur then the signal in (a) will be
appropriately reduced while that in (b) will be unaffected. On
the other hand if $\nu_e$ oscillates to a sterile state then both
the  CC and NC signals will suffer depletions.

\subsection{The variables $M_n$ and $r_n$}

One set of variables, immune to
the absolute normalisation of the initial flux, that we
propose for the extraction of oscillation signals consists of
$M_n$, the normalised $n$-th moments of the neutrino
distributions observed at SK and SNO.  Specifically,
\begin{equation}
M_n = \frac{\int N_{i}(E) E^n dE}{\int N_{i}(E) dE} 
\label{Mn}
\end{equation}
where $i$ stands for SK or SNO. It is seen from the definition
that the uncertainty in the overall normalisation of the initial
neutrino flux cancels out from $M_n$.

If neutrino oscillations are operative then we have
\begin{equation}
N_{SK}(E) =  f(E) \left\{P_{\nu_e \rightarrow
\nu_e}(E, \Delta, \vartheta) \sigma^{e}_{SK}(E) + P_{\nu_e
\rightarrow \nu_{\mu}}(E, \Delta, \vartheta)
\sigma^{\mu}_{SK}(E)\right\} \epsilon_{SK} N^0_{SK}
\label{NSK}
\end{equation}
for oscillation to any sequential neutrino, chosen to be $\nu_\mu$
in the above. Here, $f(E)$ stands for the initial Boron-neutrino
fluence, $\epsilon_{SK}$ for the detection efficiency which, for
the sake of simplicity, is assumed to be energy independent,  and
$N^0_{SK}$ for the number of electrons in the SK detector off
which the neutrinos may scatter. $\sigma^{e}_{SK}(E)$ is the
$\nu_e$ scattering cross-section with both NC and CC
contributions whereas $\sigma^{\mu}_{SK}(E)$ is the $\nu_{\mu}$
cross-section obtained from the NC interaction alone.

If the solar neutrinos oscillate to a sterile neutrino then 
eq. (\ref{NSK}) will be replaced by
\begin{equation}
N_{SK}(E) =  f(E) P_{\nu_e \rightarrow \nu_e}(E,
\Delta,\vartheta) \sigma^{e}_{SK}(E) \epsilon_{SK} N^0_{SK}
\label{NSKs}
\end{equation}

The SuperKamiokande detector uses 32 ktons of light water in
which electrons scattered by $\nu_e$ -- through both charged and
neutral current interactions -- are detected {\em via}
\u{C}erenkov radiation.  The $\nu_e - e^-$ scattering
cross-section is  $\sigma^{e}_{SK} = 9.4 \times 10^{-44} {\rm
cm}^2\, (E /10\, {\rm MeV})$ \cite {raf}. Electrons interact with
the $\nu_{\mu}$  only through neutral currents
with a cross-section $\sigma^{\mu}_{SK} = 1.6 \times
10^{-44} {\rm cm}^2\, (E /{\rm 10\, MeV})$ \cite {raf}. 

Only the CC contributions are relevant at SNO for the determination of
the spectrum and we get:
\begin{equation}
N^{c.c}_{SNO}(E) =  f(E) P_{\nu_e \rightarrow \nu_e}(E,
\Delta,\vartheta)\sigma^{c.c.}_{SNO}(E) \epsilon^{c.c.}_{SNO}
N^0_{SNO}
\label{NSNOcc}
\end{equation}
$N^0_{SNO}$ is the number of deuteron nuclei in the SNO detector
and $\epsilon^{c.c}_{SNO}$ represents the CC detection efficiency
assumed to be independent of the energy. The above result is
valid for oscillation to sequential as well as sterile neutrinos
since neither of them can interact {\em via} the charged current.

The SNO detector has 1kton of $D_2O$ and neutrinos are primarily
detected through the charged and neutral current disintegration
of the deuteron: $\nu_e + d \rightarrow e^- + p + p, \,\,
\nu + d \rightarrow \nu + p + n$, respectively. The $e^-$ in the
CC reaction is identified through its \u{C}erenkov radiation.
The neutral current process is signalled by the detection of the
neutron for which several methods are under consideration. Since
the neutral current detection is calorimetric, only the
integrated signal is measured through this channel.  The shape of
the signal is measured using the charged current process.
For this cross-section we use
$\sigma^{c.c.}_{SNO} = 1.7 \times 10^{-44}{\rm cm}^2\, (E -
1.44)^{2.3}$ where $E$ is in MeV \cite {postb}.

Some results for $M_1$ and $M_2$ are presented in Table 1 for
different values of the mass splitting $\Delta$ and the mixing
angle $\vartheta$.  Notice that at SNO, these variables cannot
distinguish between the sequential and sterile neutrino
alternatives (for a variable suited to this purpose, $R_{SNO}$,
see the following sub-section). It is seen from Table 1 that at
SuperKamiokande for the smaller mixing angle ($\vartheta =30^o$),
for any $\Delta$ the difference between the values of $M_1$ (as
well as $M_2$) for the sequential and sterile neutrino cases does
not exceed 5\% but the variation from the no oscillation ($\Delta
= 0$) limit can be as large as 10\%. Present data tend to favour
a vacuum oscillation mixing angle close to maximal and in the
$\vartheta = 45^o$ case the ranges of variation of $M_1$ and
$M_2$ are significantly larger and a distinction between the
sequential and sterile alternatives ought to be possible. At SNO,
$M_1$ and $M_2$ vary over larger ranges and, in particular, at
$\Delta = 0.6 \times 10^{-10}$ eV$^2$ and $\vartheta = 45^o$,
which are the currently indicated values, it is as much as 15\%
for $M_1$ and 20\% for $M_2$.

In order to assess the merits of these variables, it needs to
be noted first that for both the SNO CC and SK signals, what is
experimentally measured using the \u{C}erenkov technique is
the energy of the outgoing electron. In the case of SNO, the
large mass of the deuteron forces the electron to move in the
direction of the incident neutrino. Further, since the recoiling
hadrons are heavy, the electron's energy equals the incident
neutrino energy less the threshold energy for the CC reaction,
1.44 MeV. For SK there is a unique correlation between the
electron's energy and scattering angle with the neutrino energy.
Thus the neutrino spectrum can be readily reconstructed from the
measured electron energy for both experiments using the
well-known cross-sections for the appropriate scattering process.
The huge sizes of both detectors ensure that the error in the
final results will be dominated by systematic uncertainties.  A
careful analysis of the level of precision expected at these
experiments has been done in ref. \cite{blsno} and a few per cent
can be considered as a safe estimate.

In Fig. 1 we present the contours of constant $M_1$ in the
$\Delta - \vartheta$ plane for the SK sequential (1a), SK sterile
(1b), and SNO (1c) cases. The symmetry of these contours with
respect to the $\vartheta = 45^o$ line is a consequence of eq.
(\ref{ex}). Using these contours, a precise
measurement of $M_1$ will immediately constrain the parameters of
the neutrino sector. It can also be seen that the choice $\Delta
= 0.6 \times 10^{-10}$ eV$^2$ and $\vartheta = 45^o$ which is
preferred by the current data lies in a region where $M_1$
changes rather sharply. Thus $M_1$ may allow a more precise
determination of the oscillation parameters.

We have also examined the ratios
\begin{equation}
r_n = \frac{(M_n)_{SK}}{(M_n)_{SNO}}
\label{rn}
\end{equation}
Before discussing these variables we must make a cautionary
remark that the inherent risks of drawing conclusions by
combining the results from two different experiments with very
different systematics should not be underestimated.  Nonetheless,
with high statistics data from both experiments, the variables
$r_n$ may be taken as indicative of the oscillation parameters.

In Fig. 2 $r_1$, $r_2$, and $r_3$ are presented as a function of
$\Delta$ for two values of the mixing angle, $\vartheta = 45^o$
and $15^o$. Note that around the preferred $\Delta = 0.6 \times
10^{-10}$ eV$^2$ the distinction between the sequential and
sterile neutrino cases is especially pronounced for the mixing
angle of $45^o$ and the variation of $r_1$, $r_2$, and $r_3$
from the no-oscillation reference value can be larger than 25\%.

In Fig. 3 we present the contours of constant $r_2$ in the
$\Delta - \vartheta$ plane for both the sequential and sterile
neutrino cases while in Fig. 4 similar contours for $r_3$ are
shown. For both $r_2$ and $r_3$, we have presented contours for
two values of the variable larger than the no-oscillation limit
while two are smaller. Notice that these contours alternate as a
function of $\Delta$ which is a reflection of the oscillating
behaviour of $r_{2,3}$ seen in Fig.  2. The sterile and
sequential cases are especially different for the smallest values
of $\Delta$.

\subsection{The variable $R_{SNO}$}

The SNO experiment will enable separate detection of the solar
neutrinos through charged current and neutral current reactions of
which the latter will only be a calorimetric measurement. As
already noted, $\nu_\mu$ or $\nu_\tau$ produced as a result of
oscillation will register at SNO {\em via} NC interactions with full
strength but their energy will not permit charged current
interactions.  Therefore the ratio, $R_{SNO}$, of the total
signal in the NC channel, $\int N_{SNO}^{n.c.}$, to the total
(energy integrated) signal in the CC channel, $\int
N_{SNO}^{c.c.}$, is a good probe for oscillations. Thus
\begin{equation}
R_{SNO} = \frac{\int N_{SNO}^{n.c.}}{\int N_{SNO}^{c.c.}}
\label{rsno}
\end{equation}
where, for oscillation to a sequential neutrino,
\begin{equation}
\int N_{SNO}^{n.c.} = \int  f(E)
\sigma^{n.c.}_{SNO}(E) \epsilon^{n.c.}_{SNO} N^0_{SNO} dE
\label{snonc}
\end{equation}
in which $\epsilon^{n.c.}_{SNO}$ is the detection efficiency 
for the NC channel and
\begin{equation}
\int N_{SNO}^{c.c.} = \int  f(E) P_{\nu_e \rightarrow \nu_e}(E,
\Delta, \vartheta)\sigma^{c.c.}_{SNO}(E) \epsilon^{c.c}_{SNO}
N^0_{SNO} dE
\label{snocc}
\end{equation}
It is obvious that $R_{SNO}$ is independent of the absolute
normalisation of the initial neutrino flux $f(E)$ and depends
only on its shape.

If oscillations to sterile neutrinos take place then eq.
(\ref{snonc}) is replaced by:
\begin{equation}
\int N_{SNO}^{n.c.} = \int  f(E) P_{\nu_e\rightarrow \nu_e}(E,
\Delta,\vartheta) \sigma^{n.c.}_{SNO}(E) \epsilon^{n.c.}_{SNO}
N^0_{SNO} dE
\end{equation}
and eq. (\ref{snocc}) is unchanged.

For the NC cross-section $\sigma^{n.c.}_{SNO}$ we use  $0.85
\times 10^{-44}{\rm cm}^2\, (E - 2.2)^{2.3}$ where $E$ is in MeV
\cite {postb}.  For simplicity, we have assumed
$\epsilon^{n.c.}_{SNO}$ to be independent of the energy and
further equal to the efficiency of the CC reaction
$\epsilon^{c.c.}_{SNO}$. If instead,
$\epsilon^{n.c.}_{SNO}/\epsilon^{c.c.}_{SNO} = r_{\epsilon}$ and
it can be taken to be independent of the energy to a good
approximation, then our results for $R_{SNO}$ will be multiplied
by this factor.  The predictions can be improved by using more
refined expressions for the NC and CC cross-sections and more
realistic inputs for detection efficiencies. Our purpose is to
illustrate the utility of the proposed variables and we refrain
from this exercise here.

Results for $R_{SNO}$ for different values of the mass splitting
$\Delta$ are presented in Table 2 for three choices of the mixing
angle $\vartheta = 15^o, \;\; 30^o$ and $45^o$. In the absence of
oscillations we find $R_{SNO}$ = 0.382.  Oscillation to
sequential neutrinos decreases the denominator of eq.
(\ref{rsno}) while the numerator is unaffected. Thus $R_{SNO}$
increases, very prominently for larger mixing angles, in this
scenario. For sterile neutrinos both the NC and CC contributions
are affected and we find little change in $R_{SNO}$.  Thus,
$R_{SNO}$ will not be able to yield much information if
oscillation to a sterile state is operative. A significantly
different $R_{SNO}$ from its no-oscillation limit will be  a
clear indication of oscillation to a sequential neutrino. In
particular, for $\Delta = 0.6 \times 10^{-10}$ eV$^2$ and
$\vartheta = 45^o$, $R_{SNO}$ will be as high as 2.1 for
oscillation to sequential neutrinos.

In Fig. 5, we present contours of constant $R_{SNO}$ in
the $\Delta$-$\vartheta$ plane for oscillation to sequential
neutrinos. The symmetry of the contours about $\vartheta = 45^o$
is expected. At $\Delta = 0$ or $\vartheta = 0^o$ or $90^o$ the
limit of no oscillations will be obtained.  Values of $R_{SNO}$
as high as 0.99 can only be achieved for smaller values of
$\Delta$.

\section{Supernova neutrinos and the new variables}

SuperKamiokande and SNO will also serve as telescopes for
neutrinos from supernova explosions. The detection of the SN1987A
neutrinos gives us confidence that in the event of a supernova
explosion both facilities will observe signals which will be
pointing in the same direction and arrive at the same time.  Here
again, as for solar neutrinos, the proposed variables may be used
to look for neutrino oscillations. The characteristic energy of
the neutrinos being of the order of 10 MeV and the huge distance
scales involved (several kpc) put the supernova in a unique
position as a probe for oscillations governed by mass scales as
small as $10^{-18}-10^{-19}$ eV$^2$. We consider collapse phase
and post-bounce era neutrinos in turn.

\subsection{Collapse phase neutrinos}

In the precursor to a supernova explosion the core of a massive
star collapses under its own gravity once the nuclear burning
stops and the pressure support from degenerate electrons is
reduced due to electron capture. In this collapse phase, a burst
of $\nu_e$, produced due to electron capture, is emitted in a
timescale of about ten milliseconds. The burst stops when the
density of the star becomes so large that neutrinos get trapped
within. Though the number of neutrinos emitted in this phase is
much less than that in the post-bounce era (see later), it has the
advantage of being  a pure electron neutrino beam. In this
subsection we examine these neutrinos in the light of the new
variables. The shape of the supernova neutrino spectra
cannot be predicted with the precision of the solar case. We wish
to establish that, nonetheless, the variables under discussion
turn out to be useful in the search for neutrino oscillations.

In the event of  a supernova explosion occuring sufficiently
nearby (typically 1 kpc away) SK and SNO should detect a
substantial number of neutrinos from the stellar core collapse
phase \cite{sr,fks}.  The collapse phase neutrino spectra from a
realistic range of nuclear physics inputs as well as several
stellar masses on the main sequence are presented in refs.
\cite{sr,fks} and in the rest of our discussion in this
subsection we use the neutrino spectrum presented there for a
15M$_\odot$ star at a distance of 1 kpc. If the supernova is much
further away then the flux of neutrinos will be too weak to be
detected with significant statistics.

We can readily adapt eqs.  (\ref{Mn}--\ref{NSNOcc}) to calculate
the variables $M_n$ for neutrinos from the collapse phase. It
needs to be mentioned  that supernova neutrinos have higher
energy than in the solar case and small additional contributions
to the SK and SNO signals will come from the CC and NC processes
$\nu_e + ^{16}O \rightarrow e^- + ^{16}F$ and $\nu_x + ^{16}O
\rightarrow \nu_x + \gamma + X$ respectively which have
thresholds of 15.4 MeV and 15.0 MeV. We have ignored these
contributions and estimate that in this case their effect will
not exceed about 10\% of the signal.

$M_1$ and $M_2$ for different values of $\Delta$ are presented in
Table 3 for maximal mixing ($\vartheta = 45^o$).  Results for
oscillation to sequential and sterile neutrinos are separately
presented for SK.  As noted for solar neutrinos, the SNO CC
signal does not distinguish between these alternatives.  The
ratios $r_1$ and $r_2$ (see eq. (\ref {rn})) are also presented
in Table 3. It is seen from this Table that the variation in
$M_1$ can be 15\%  at both SK and SNO for some values of the
oscillation parameters while for $M_2$ it can be as large as 40\%
(25\%) at SNO (SK). At SK, the variation is larger in the sterile
neutrino scenario. 

As mentioned earlier,  unlike the solar neutrino case, the shape
of the initial collapse phase neutrino spectrum is not precisely
known and the above variation has to be set against the
uncertainty in these variables which may arise, even in the
absence of oscillations, from the imperfect knowledge of the
initial neutrino flux.  The neutrino energy spectrum receives
contributions from electron capture on both free protons and
heavy nuclei (in the {\em fp} shell). In ref. \cite{sr,fks} the
abundance distribution of these are self-consistently determined
with the evolution of thermodynamic conditions as collapse
proceeds. As extreme conservative limits of the uncertainty in
the neutrino spectrum we consider the situations where in one
case the electron capture is only on free protons while in the
other it is exclusively on heavy nuclei.  Further, we estimate
the uncertainties due to the unknown mass of the progenitor by
considering a 15M$_\odot$ as well as a 25M$_\odot$ star.  The
results for $M_1$, $M_2$ as also $r_1$, and $r_2$  for the above
possibilities are presented in Table 4.  It is seen from this
Table that, in fact, the considered variation of the initial
neutrino spectrum can lead to uncertainties in $M_1$ and $M_2$
larger than that due to oscillations which were presented in
Table 3.  Thus, for the collapse phase neutrinos, $M_1$ and $M_2$
cannot be used to unequivocally signal neutrino oscillations. The
situation is somewhat better with the variables $r_1$ and $r_2$
in the sense that variation due to uncertainties in the initial
spectrum can be at most 0.83 - 0.91 and 0.72 - 0.84 respectively
and cannot mask an effect due to oscillations at least for some
ranges of the mixing parameters. In Fig.  6, we present $r_1$,
$r_2$, and $r_3$ as a function of $\Delta $ for two values of the
mixing angle $\vartheta$. Notice that around $\Delta = 1 \times
10^{-18}$ eV$^2$ these variables can clearly distinguish between
oscillation to a sequential or to a sterile neutrino for
$\vartheta = 45^o$.

The variable $R_{SNO}$ -- eq. (\ref{rsno}) -- turns out to be
very efficient to look for oscillations in the collapse phase
neutrino data. The variation of $R_{SNO}$ with the mass splitting
$\Delta$  is shown in Table 5 where the mixing angle has been
chosen to be $\vartheta = 45^o$. From these results it is seen
that $R_{SNO}$ can be increased several times if oscillation to
sequential neutrinos occur and can achieve values as high as 1.7.
It should be borne in mind that these results are for maximal
mixing and there will be a suppression for other mixing angles.
The sterile neutrino alternative affects the variable only
marginally.  The significance of the results presented in Table 5
can only be gauged once a comparison is made with the uncertainty
in $R_{SNO}$ due to the imprecise knowledge of the initial
neutrino flux.  In Table 6 results are presented for $R_{SNO}$ in
the absence of oscillations for both $15M_{\odot}$ and
$25M_{\odot}$ stars.  It is seen from this Table that the value
of $R_{SNO}$ is not sensitive to the typical examples of stellar
collapse considered, which include combinations of initial
conditions as reflected in the zero age main sequence mass of the
pre-supernova star, matrix elements of the electron capture on
heavy nuclei etc., and varies within the range 0.382 -- 0.445.
Therefore, if $R_{SNO}$ is found to be very different from the
predicted no-oscillation value then this difference cannot be
attributed to the range of variations expected from astrophysical
and nuclear physics grounds and will point towards oscillation to
a sequential neutrino.  In Fig. 7, we present contours of
constant $R_{SNO}$ in the $\Delta - \vartheta$ plane for collapse
phase neutrinos for oscillation to a sequential state. Notice
that the highest values of $R_{SNO}$ can be achieved only for
smaller choices of $\Delta$.

\subsection{Post-bounce epoch neutrinos}

In the post-bounce epoch, thermal neutrinos and also
anti-neutrinos of all three flavours are emitted on a time scale
of a few seconds.  The flux of these neutrinos is higher than
those from the collapse phase and, indeed, is intense enough that
neutrinos from this  era of a supernova event at a distance of 10
kpc will register a statistically significant signal at SK and
SNO. For example, the SN1987A explosion registered 11 and 8
events respectively at the smaller Kamiokande and IMB detectors.
The sequence of arrival-times of these neutrinos can yield
information on neutrino masses \cite{tot1}. In contrast, the
discussion here, based on neutrino oscillations, will shed light
on neutrino mass splittings.

The $\tau$- and $\mu$-type neutrinos and anti-neutrinos emitted
during this era are pair-produced in the supernova and, in the
absence of degeneracies, are predicted to have the same energy
spectrum. The $\nu_e$ and $\bar{\nu}_e$ have different spectra
and due to their additional charged current interactions inside
the star have lower energy than the $\nu_\mu$ and $\nu_\tau$.
Neutrino oscillations between these states will therefore induce
a complicated energy distribution of $\nu$ and $\bar{\nu}$ of
different flavours \cite{cmk}. In this sub-section we examine how
the spectral shape variables, introduced earlier, can be utilised
to discern oscillations in such a signal.

In the following analysis three different post-bounce neutrino
spectra are at play; namely, those for the $\nu_e$, the
$\bar{\nu}_e$ and the $\nu_x$ where the latter stands for
neutrinos as well as anti-neutrinos of the $\mu$ and $\tau$
types. Here, for the purposes of illustration, we use the spectra
extracted from the results presented in \cite{tot2}. We restrict
ourselves, as earlier, to two flavour vacuum neutrino
oscillations. Since the $\tau$- and $\mu$-flavours are on an
equal footing as far as the spectra and the detectors are
concerned, we can consider mixing with any one of these as
characteristic of oscillation to a sequential neutrino. Thus, for
example, if we consider $\nu_e \leftrightarrow \nu_\mu$
oscillations, then the $\nu_\tau$ neutrinos and anti-neutrinos
will be entirely unaffected while both neutrinos and
anti-neutrinos of the electron and muon type undergo
oscillations.  Had we considered MSW resonant flavour conversion
instead, then, depending on the sign of the mass squared
difference, either the neutrinos or the anti-neutrinos would have
undergone conversion.

Unlike the previously discussed solar and collapse phase
neutrinos, here, for the first time, we have both neutrinos and
anti-neutrinos of all three flavours in the initial beam.  This
adds several new features to the analysis. For example, while
oscillations to sequential or sterile neutrinos affected the CC
signal in the same manner in the earlier cases, this is no longer
the case. Thus, for post-bounce neutrinos, oscillation to
$\nu_\mu$ results in some of the $\nu_e$ changing to muon
neutrinos but at the same time some electron neutrinos are
produced from the $\nu_\mu$ in the original beam. Since there are
no sterile neutrinos initially, the situation will be different
if the $\nu_e$ oscillates to a sterile state. Thus even {\em via}
the CC interactions the two cases can be distinguished.  Further
at SNO, in addition to the NC interactions, the $\bar{\nu}_e$
will register via the CC reaction $\bar{\nu}_e + d
\rightarrow e^+ + n + n$ which has a threshold of 4.03 MeV
\cite{postb}.  We have not included a small contribution from the
process $\bar{\nu}_e + ^{16}O \rightarrow e^+ + ^{16}N$.

If we indicate the time integrated energy spectra of $\nu_e$,
$\bar{\nu}_e$ and the neutrinos (and anti-neutrinos) of the $\mu$
and $\tau$ flavours by $f^e(E)$, $\bar{f}^e(E)$ and $f^x(E)$
respectively then the observed signal at SK can be written as:
\begin{eqnarray}
N_{SK}(E) &=&  \left[\left\{ f^e(E) P_{\nu_e \rightarrow
\nu_e}(E, \Delta, \vartheta) + f^x(E) P_{\nu_\mu \rightarrow
\nu_e}(E, \Delta, \vartheta) \right\} \sigma^{e}_{SK}(E) +
\right. \nonumber \\
&& \left. \left\{f^{e}(E) P_{\nu_e \rightarrow
\nu_\mu}(E, \Delta, \vartheta) + f^x(E) P_{\nu_\mu \rightarrow
\nu_\mu}(E, \Delta, \vartheta) + f^x(E) \right\}
\sigma^{\mu}_{SK}(E) + \right. \nonumber \\
&&\left.\left\{ \bar{f}^e(E) P_{\nu_e \rightarrow
\nu_e}(E, \Delta, \vartheta) + f^x(E) P_{\nu_\mu \rightarrow
\nu_e}(E, \Delta, \vartheta)  \right\}
\bar{\sigma}^{e}_{SK}(E) + \right. \nonumber \\ && \left.
\left\{\bar{f}^{e}(E) P_{\nu_e \rightarrow
\nu_\mu}(E, \Delta, \vartheta) + f^x(E) P_{\nu_\mu \rightarrow
\nu_\mu}(E, \Delta, \vartheta) + f^x(E) \right\}
\bar{\sigma}^{\mu}_{SK}(E) \right] \epsilon_{SK} N^0_{SK} \nonumber \\
\label{NSKpb}
\end{eqnarray}
for oscillation to any sequential neutrino, chosen to be $\nu_\mu$
in the above. Here $\bar{\sigma}^x_{SK}(E)$ is the
$\bar{\nu}_\mu$ or $\bar{\nu}_\tau$ scattering cross-section off
electrons which proceeds {\em via} the neutral current and is
$1.3 \times 10^{-44}\, {\rm cm}^2\, E/(10\, {\rm MeV})$. For the
$\bar{\nu}_e$, there is a a charged current contribution so that
the total $\bar{\nu}_e-e$ scattering cross-section is $3.9\times
10^{-44} {\rm cm}^2\, E/(10\, {\rm MeV})$.
$\bar{\sigma}^e_{SK}(E)$ receives an additional (dominant)
contribution from the process $\bar{\nu}_e + p
\rightarrow e^+ + n$ which is $9.4 \times 10^{-42} {\rm cm}^2\, p_e
E_e/(10\, {\rm MeV})^2$ where $p_e$ is the electron momentum and
$E_e = E - 1.3$ MeV its energy \cite{raf}.

If instead, oscillations to a sterile neutrino are operative, then 
eq. (\ref{NSKpb}) will be replaced by
\begin{eqnarray}
N_{SK}(E) &=&  \left[f^e(E) P_{\nu_e \rightarrow
\nu_e}(E, \Delta, \vartheta)  \sigma^{e}_{SK}(E) 
+ 2 f^x(E) \sigma^{\mu}_{SK}(E) \right. \nonumber \\
&& \left. + \bar{f}^e(E) P_{\nu_e \rightarrow
\nu_e}(E, \Delta, \vartheta) 
\bar{\sigma}^e_{SK}(E) + 2 f^x(E) \bar{\sigma}^x_{SK}(E) \right]
\epsilon_{SK} N^0_{SK} 
\label{NSKspb}
\end{eqnarray}

As discussed earlier, only the CC contributions are relevant at
SNO for the extraction of the spectral shape and in this case the
relevant formula valid for oscillation to a sequential state is:
\begin{eqnarray}
N^{c.c}_{SNO}(E) &=&  \left[ \left\{f^e(E) P_{\nu_e \rightarrow
\nu_e}(E,
\Delta,\vartheta) + f^x(E) P_{\nu_\mu \rightarrow \nu_e}(E,
\Delta,\vartheta) \right\}\sigma^{c.c.}_{SNO}(E) + \right. \nonumber \\
&& \left. \left\{\bar{f}^e(E) P_{\nu_e \rightarrow \nu_e}(E,
\Delta,\vartheta) + f^x(E) P_{\nu_\mu \rightarrow \nu_e}(E,
\Delta,\vartheta) \right\}\bar{\sigma}^{c.c.}_{SNO}(E) \right]
\epsilon^{c.c.}_{SNO} N^0_{SNO} \nonumber \\
\label{NSNcpb}
\end{eqnarray}
while for oscillation to sterile neutrinos it is:
\begin{eqnarray}
N^{c.c}_{SNO}(E) &=&  \left[ f^e(E) P_{\nu_e \rightarrow
\nu_e}(E,\Delta,\vartheta) \sigma^{c.c.}_{SNO}(E) + \right.
\nonumber \\  &&  \left. \bar{f}^e(E) P_{\nu_e \rightarrow
\nu_e}(E,\Delta,\vartheta) \bar{\sigma}^{c.c.}_{SNO}(E) \right]
\epsilon^{c.c.}_{SNO} N^0_{SNO}
\label{NSNcspb}
\end{eqnarray}

In the above, we have only considered two flavour neutrino
oscillations that involve the electron neutrino. This is because
the SNO and SK detectors are primarily geared to look for the
$\nu_e$. If two flavour oscillation occurs between the $\nu_\mu
\leftrightarrow \nu_\tau$ states then neither the SNO nor the SK
signal will be affected at all. If $\nu_\mu$ or $\nu_\tau$
oscillates to a sterile state then though the SNO signal remains
unchanged,  there will, indeed, be a small depletion in the SK
signal since the latter lacks the NC interaction of the
sequential neutrino. We have not discussed this case since the
effect will be small.

Some results for $M_1$ and $M_2$ for post-bounce epoch neutrinos
are presented in Table 7 for different values of the mass
splitting $\Delta$ for the mixing angle $\vartheta = 45^o$.  One
major difference in this Table from those of the situations
discussed earlier is that for post-bounce neutrinos SNO can
distinguish between oscillation to sequential and sterile
neutrinos. As noted earlier, this is because the post-bounce
epoch beam has $\nu_\mu$ and $\nu_\tau$ components in addition to
electron neutrinos.  Since the spectra of the electron- and
muon-type neutrinos are different, the net effect of oscillations
is to change the CC signal by an amount different from that for
the case of oscillation to sterile neutrinos. Due to the higher
energy of the $\nu_\mu$ and $\nu_\tau$, oscillation to sequential
neutrinos always increases the signal at both SK and SNO while
for the sterile case both larger and smaller values are possible
depending on $\Delta$. It is seen from Table 7 that the effect of
oscillations is most pronounced around $\Delta = 1.2
\times 10^{-19}$ eV$^2$. For example, for the sterile
alternative, for $M_1$ the deviation from the no-oscillation
value is about 25\% (20\%) at SK (SNO) for $M_2$ it is 45\%
(39\%).

In order to gauge the utility of these variables, it needs to be
first ascertained to what extent they are sensitive to changes in
the input spectra of the neutrinos. These spectra are extracted
from results on the evolution of supernova explosions and it is
not easy to assess the range of uncertainty, {\em ab initio}. We
consider as a conservative upper limit a variation of $\pm 30\%$
in the absolute normalisation of each of $f^e(E)$,
$\bar{f}^e(E)$, and $f^x(E)$. In Table 8 are presented results
for $M_{1,2}$ and $r_{1,2}$, in the absence of oscillations,
where independent variation of each spectrum in this range takes
place. Notice that the variables are remarkably stable.  Hence
for neutrinos from the post-bounce epoch, the variables $M_1$,
$M_2$, $r_1$, and $r_2$ can be powerful tools to probe for
oscillations.

In Fig. 8 $r_1$, $r_2$, and $r_3$ are presented as a function of
$\Delta$ for two values of the mixing angle, $\vartheta = 45^o$
and $15^o$. Notice that for oscillation to sequential neutrinos
the variables are always smaller than the no oscillation limit
while in the sterile case both larger and smaller values are
possible. This difference can be attributed to the presence of
the other sequential neutrinos in the parent beam.

In Fig. 9 we present the contours of constant $r_2$ in the
$\Delta - \vartheta$ plane.   For oscillation to  sequential
neutrinos, we have presented contours for four values of $r_2$
which are all less than the no-oscillation limit as dictated by
Fig. 8. In the sterile case, however, we have presented contours
for two values of the variable larger than the no-oscillation
limit while two are smaller. The vastly different nature of the
contours for the sequential and sterile alternatives underscore
the utility of these variables to pin-point the kind of
oscillation at work.

For post-bounce epoch neutrinos the ratio, $R_{SNO}$, is also an
effective probe for oscillations. In contrast, to the case of
solar neutrinos, here it turns out to be sensitive to
oscillations of an electron neutrino to a sterile state. As
before,
\begin{equation}
R_{SNO} = \frac{\int N_{SNO}^{n.c.}}{\int N_{SNO}^{c.c.}}
\label{rsnopb}
\end{equation}
where for oscillation to a sequential neutrino
\begin{eqnarray}
\int N_{SNO}^{n.c.} = && \int \left[ \left\{f^e(E) + 2
f^x(E)\right\}\sigma^{n.c.}_{SNO}(E) \right. \nonumber \\
&& \left. + \left\{\bar{f}^e(E) + 2
f^x(E)\right\}\bar{\sigma}^{n.c.}_{SNO}(E) \right]
\epsilon^{n.c.}_{SNO} N^0_{SNO} dE
\label{snoncpb}
\end{eqnarray}
where to a good approximation the NC cross-section for
anti-neutrinos of all flavours $\bar{\sigma}^{n.c.}_{SNO}(E) =
\sigma^{n.c.}_{SNO}(E)$ \cite{postb}. For the sterile neutrino
alternative
\begin{eqnarray}
\int N_{SNO}^{n.c.} = && \int \left[\left\{f^e(E) P_{\nu_e
\rightarrow \nu_e}(E, \Delta, \vartheta) + 2
f^x(E)\right\}\sigma^{n.c.}_{SNO}(E) \right. \nonumber \\
&& \left. + \left\{\bar{f}^e(E)
P_{\nu_e \rightarrow \nu_e}(E, \Delta, \vartheta) + 2
f^x(E)\right\}\bar{\sigma}^{n.c.}_{SNO}(E) \right]
\epsilon^{n.c.}_{SNO} N^0_{SNO} dE
\end{eqnarray}
and
\begin{equation}
\int N_{SNO}^{c.c.} =  \int N^{c.c}_{SNO}(E) dE
\end{equation}
where $N^{c.c}_{SNO}(E)$ is given by eq. (\ref{NSNcpb}) or eq.
(\ref{NSNcspb}) depending on whether oscillation of electron
neutrinos takes place to sequential or sterile neutrinos,
respectively. In Table 9, we present $R_{SNO}$ as a function of 
$\Delta$ for  $\vartheta = 15^o$ and $45^o$. It is evident from
this Table that $R_{SNO}$ varies over a wide range (more than
100\% of the reference no-oscillation value) in the sterile
neutrino alternative while for oscillation to sequential neutrinos
it is much less. Further, in the former case the effect of
neutrino oscillations is always to increase $R_{SNO}$ while for
the latter the effect is in the opposite direction
(barring  for very small $\Delta$).

To judge to what extent $R_{SNO}$ is sensitive, in the absence of
oscillations, to the initial neutrino spectra we have
conservatively let these vary by $\pm$ 30\%. These results, along
with those where the variation is restricted to $\pm$ 10\%, are
presented in Table 10. Notice that a $\pm $ 10\% variation
in the initial flux can lead to values of $R_{SNO}$ which can be
achieved by oscillation to sequential neutrinos. On the other
hand the range of variation obtained in the sterile neutrino
alternative cannot be covered  by  even a $\pm$ 30\% change in
the intial flux. Therefore, we conclude that for the post-bounce
epoch neutrinos $R_{SNO}$ is a useful diagnostic tool only for
oscillation to a sterile neutrino. In Fig. 10 are presented the
contours  of constant $R_{SNO}$ in the $\Delta - \vartheta$ plane
for oscillation of the $\nu_e$ to a sterile state.  Note that the
largest values of $R_{SNO}$ can be achieved only for the small
$\Delta $ region.

\section{Discussions and Conclusions}

In this work we have elaborated on several variables that probe
the shape of neutrino spectra seen at SK and SNO in a manner
independent of the absolute normalisation of the neutrino
fluence. The variables can be fruitfully used since the data from
these experiments will be of unprecedented high statistics. As
such they are useful to detect the modification of a neutrino
spectrum by any process. Though, in this paper, we have
illustrated their utility to signal two flavour vacuum neutrino
oscillations only, similar analyses can be readily carried out
for MSW resonant flavour conversion, multi-generational mixing,
spin precession in a magnetic field, neutrino decay, {\em etc.}.
We hope to return to these issues in subsequent work.

One class of variables that was proposed in this work, $M_n$, are
the normalised $n$-th moments of the observed neutrino spectra at
SK and SNO and their ratios, $r_n = (M_n)_{SK}/(M_n)_{SNO}$.
Another variable discussed in this paper, $R_{SNO}$, specific to
SNO, is the ratio of the energy integrated NC signal to the
energy integrated CC signal.

These variables are most appropriate for solar neutrinos.
SuperKamiokande and SNO will be sensitive only to the so-called
$^8B$ neutrinos from the sun, the shape of whose energy spectrum is known
precisely but the absolute normalisation is comparatively much
less certain. (We have ignored a small contribution from
$hep$ neutrinos.) All the variables discussed in this work are
independent of this absolute normalisation. We have found that  $M_1$
and $M_2$ for the solar neutrinos will allow a distinction
between the alternatives of oscillation of the $\nu_e$ to
sequential and sterile neutrinos. The related variables $r_1$,
$r_2$, and $r_3$ can also be conveniently used to signal
oscillations and distinguish between sequential and sterile
neutrinos. Since oscillation to sterile neutrinos affect both the
NC and CC signals at SNO in similar manners, $R_{SNO}$ is rather
insensitive in this case. It will be useful to detect oscillation
to sequential neutrinos.
 
In addition to solar neutrinos, SK and SNO will also serve as neutrino
telescopes for supernova explosions. The mass splitting $\Delta$
that can be explored {\em via} supernova neutrinos $\sim
10^{-18}$ or $10^{-19}$ eV$^2$, is, indeed, very tiny.  The
energy and length scales associated with supernova neutrinos
provide a unique window for very small mass splittings -- a point
noted earlier in ref. \cite {rs}.  It has been speculated
that oscillation of neutrinos from Active Galactic Nuclei or
Gamma Ray Bursts will also be sensitive to such small $\Delta $
\cite {hal}.  

Neutrinos are emitted at two stages of a supernova explosion.
Though the earlier collapse phase neutrinos have the advantage of
being a pure $\nu_e$ beam, their flux is weaker and such an event
will be detectable only if it occurs within a distance of 1 kpc.
Unlike the solar case, there is also some uncertainty in the
spectrum of the emitted neutrinos. We have found that due to
these uncertainties it is not possible to unequivocally signal
neutrino oscillations {\em via} $M_1$ and $M_2$. However,
$R_{SNO}$ turns out to be a useful tool even in this case.

For the later post-bounce epoch, neutrinos and anti-neutrinos of
all three flavours are emitted. In this case, we find that the
variables $M_1$, $M_2$, $r_1$, and $r_2$ are all suitable for probing
neutrino oscillations. In the absence of oscillations, a
variation of the initial spectra by as much as $\pm$30\% is
reflected by only a few per cent change in these parameters. In
the post-bounce neutrino case, $R_{SNO}$ turns out to be useful
only for signalling oscillation to sterile neutrinos.

We conclude that the variables discussed in this work can be powerful
diagnostic tools to search for neutrino oscillations in solar and
supernova neutrino data obtained at SK and SNO. The results presented
here can be further sharpened by simulating the detector geometries,
acceptances, and detection efficiencies. These variables can also
be used for other detectors -- {\em e.g.}  ICARUS which is 
sensitive only to $^8B$ neutrinos from the sun -- which are in the
development stage.

\parindent 0pt

{\large{\bf {Acknowledgements}}}

The authors are grateful to Sandhya Choubey for her help.
This work is partially supported by the Eastern Centre for
Research in Astrophysics, India. A.R. also acknowledges a
research grant from the Council of Scientific and Industrial
Research, India.

\newpage

\parindent 0pt

Table 1: $M_1$ and $M_2$ for solar neutrinos for different values
of the mass splitting $\Delta$ for the SuperKamiokande and SNO
detectors.  For the former, results are presented for oscillation
to sequential as well as sterile neutrinos. For SNO the two cases
yield the same value of $M_n$. Two choices of the mixing angle
$\vartheta = 30^o$ and $45^o$ have been considered.

\begin{center}
\begin{tabular}{|c|c|c|c|c|c|c|c|c|c|c|c|c|}
\hline
$\Delta $ & \multicolumn{6}{|c|}{$M_1$} 
& \multicolumn{6}{|c|}{$M_2$} \\
\cline{2-13}
in & \multicolumn{3}{|c|}{$\vartheta=30^0$} & 
\multicolumn{3}{c|}{$\vartheta=45^0$} & 
\multicolumn{3}{c|}{$\vartheta=30^0$} & 
\multicolumn{3}{c|}{$\vartheta=45^0$} \\ 
\cline{2-13} 
$10^{-10}$ & \multicolumn{2}{c|} {SK} & SNO & \multicolumn{2}{c|}
{SK} & SNO &
\multicolumn{2}{c|} {SK} & SNO & \multicolumn{2}{c|} {SK} & SNO \\
\cline{2-3} \cline{5-6} \cline{8-9} \cline{11-12}
eV$^2$& seq. & st. & & seq. & st. & 
& seq. & st. & & seq. & st. & \\
\hline 
  .0&    8.49&    8.49&    9.35&    8.49&    8.49&    9.35&
76.3&   76.3&   91.4&   76.3&   76.3&   91.4 \\
  .3&    8.76&    8.83&    9.62&    8.89&    9.03&    9.76&
80.8&   82.1&   96.3&   83.0&   85.3&   98.8 \\
  .6&    8.73&    8.87&    9.81&    9.01&    9.77&   10.67&
80.9&   83.6&  100.5&   86.31&  101.0&  117.5 \\
  .9&    7.85&    7.57&    8.52&    7.34&    6.57&    7.09&
65.7&   61.2&   77.4&   57.4&   44.8&   53.1 \\
 1.2&    8.13&    8.02&    8.67&    7.94&    7.74&    8.21&
69.3&   67.2&   78.4&   65.7&   61.9&   69.4 \\
 1.5&    8.71&    8.78&    9.36&    8.82&    8.93&    9.37&
79.2&   80.0&   90.6&   80.6&   82.0&   90.1 \\
 1.8&    8.79&    8.88&    9.75&    8.94&    9.10&    9.95&
81.8&   83.5&   98.6&   84.6&   87.5&  102.2 \\
 2.1&    8.66&    8.71&    9.76&    8.75&    8.85&   10.02&
80.0&   81.2&  100.1&   82.1&   84.5&  105.5 \\
 2.4&    8.47&    8.46&    9.45&    8.46&    8.44&    9.52&
76.3&   76.3&   94.4&   76.3&   76.4&   96.7 \\
 2.7&    8.26&    8.18&    9.02&    8.12&    7.95&    8.74&
72.2&   70.7&   85.6&   69.6&   66.6&   80.8 \\
 3.0&    8.34&    8.29&    8.99&    8.25&    8.15&    8.72&
73.1&   72.0&   84.1&   71.3&   69.1&   78.6 \\
 3.5&    8.54&    8.56&    9.37&    8.57&    8.59&    9.38&
77.0&   77.2&   91.2&   77.4&   77.8&   91.1 \\
 4.0&    8.61&    8.65&    9.56&    8.67&    8.75&    9.69&
78.6&   79.3&   95.8&   79.9&   81.3&   98.5 \\
 4.5&    8.45&    8.44&    9.35&    8.43&    8.40&    9.36&
75.8&   75.6&   92.1&   75.4&   75.1&   92.5 \\
 5.0&    8.45&    8.44&    9.23&    8.43&    8.41&    9.14&
75.4&   75.1&   88.9&   74.8&   74.2&   87.1 \\
 5.5&    8.48&    8.47&    9.29&    8.47&    8.46&    9.26&
75.9&   75.8&   90.1&   75.7&   75.4&   89.2 \\
 6.0&    8.53&    8.54&    9.41&    8.55&    8.57&    9.46&
77.0&   77.3&   92.8&   77.4&   77.9&   93.7 \\
\hline 
\end{tabular}
\end{center}

\newpage

Table 2: $R_{SNO}$ for solar neutrinos for different
values of the mixing angle, $\vartheta$, and the mass splitting,
$\Delta$. Results are presented for oscillation to sequential
as well as sterile neutrinos.

\begin{center}
\begin{tabular}{|c|c|c|c|c|c|c|}
\hline
$\Delta $ & \multicolumn{6}{|c|}{$R_{SNO}$} \\
\cline{2-7}
in & \multicolumn{2}{|c|}{$\vartheta=15^0$} & 
\multicolumn{2}{c|}{$\vartheta=30^0$} & 
\multicolumn{2}{c|}{$\vartheta=45^0$} \\
\cline{2-7} 
$10^{-10}$ eV$^2$& Sequential & Sterile & Sequential & Sterile &
Sequential & Sterile \\
\hline

 0.0 &  0.382 &  0.382 &  0.382 &  0.382 & 0.382 &  0.382 \\
 0.3 &  0.422 &  0.384 &  0.532 &  0.389 & 0.613 &  0.392 \\
 0.6 &  0.480 &  0.383 &  0.991 &  0.387 & 2.117 &  0.396 \\
 0.9 &  0.467 &  0.378 &  0.848 &  0.362 & 1.428 &  0.337 \\
 1.2 &  0.438 &  0.380 &  0.623 &  0.375 & 0.788 &  0.370 \\
 1.5 &  0.422 &  0.383 &  0.537 &  0.387 & 0.620 &  0.390 \\
 1.8 &  0.417 &  0.383 &  0.512 &  0.386 & 0.577 &  0.388 \\
 2.1 &  0.431 &  0.383 &  0.582 &  0.387 & 0.706 &  0.390 \\
 2.4 &  0.444 &  0.382 &  0.660 &  0.383 & 0.873 &  0.384 \\
 2.7 &  0.444 &  0.380 &  0.658 &  0.375 & 0.867 &  0.370 \\
 3.0 &  0.444 &  0.381 &  0.659 &  0.379 & 0.869 &  0.377 \\
 3.5 &  0.431 &  0.382 &  0.582 &  0.381 & 0.705 &  0.380 \\
 4.0 &  0.434 &  0.383 &  0.597 &  0.386 & 0.735 &  0.388 \\
 4.5 &  0.435 &  0.382 &  0.606 &  0.381 & 0.753 &  0.380 \\
 5.0 &  0.440 &  0.382 &  0.634 &  0.381 & 0.813 &  0.381 \\
 5.5 &  0.437 &  0.382 &  0.614 &  0.381 & 0.770 &  0.381 \\
 6.0 &  0.434 &  0.382 &  0.597 &  0.382 & 0.735 &  0.382 \\
\hline
\end{tabular}
\end{center}

\newpage

Table 3: $M_1$, and $M_2$ for collapse phase neutrinos for
different values of the mass splitting $\Delta$ for the SK and
SNO detectors.  For the former, results are presented for
oscillation to sequential as well as sterile neutrinos. For SNO,
the two cases yield the same value of $M_n$. The ratios $r_1$ and
$r_2$ are also presented. The mixing angle $\vartheta$ has been
chosen to be $45^o$.

\begin{center}
\begin{tabular}{|c|c|c|c|c|c|c|c|c|c|c|}
\hline
$\Delta$ in & \multicolumn{3}{|c|}{$M_1$} & \multicolumn{2}{|c|}{$r_1$} 
& \multicolumn{3}{|c|}{$M_2$} & \multicolumn{2}{|c|}{$r_2$}\\
\cline{2-11}
$10^{-18}$ & \multicolumn{2}{c|} {SK} & SNO & seq. & st. &
\multicolumn{2}{c|} {SK} & SNO & seq. & st. \\
\cline{2-3} \cline{7-8} 
eV$^2$& seq. & st. & & & & seq. & st. & & &  \\
\hline 
  .0&   13.3&   13.3&   16.1& 0.83 & 0.83 & 202.4&
202.4&  281.7&  0.72 & 0.72 \\
  .6&   11.3&    9.9&   13.9& 0.81 & 0.71 &
152.3&  120.6&  234.0 & 0.65 & 0.52 \\
 1.2&   13.5&   13.6&   15.6& 0.87 & 0.87&
 203.1&  203.3&  256.5 & 0.79 & 0.79\\
 1.8&   13.9&   14.2&   17.5&  0.79 & 0.81&
 223.9&  232.3&  328.8 & 0.68 & 0.71 \\
 2.4&   12.7&   12.4&   14.9& 0.85 & 0.83&
 182.0&  172.9&  243.2 & 0.75 & 0.71\\
 3.0&   13.5&   13.5&   16.0& 0.84 & 0.84&
 204.2&  205.0&  272.8 & 0.75 & 0.75 \\
 4.0&   13.3&   13.3&   16.1& 0.83 & 0.83&
 201.1&  200.1&  284.6 & 0.71 & 0.70\\
 5.0&   13.4&   13.4&   16.1&  0.83 & 0.83&
 203.3&  203.7&  280.3 & 0.73 & 0.73\\
 6.0&   13.3&   13.3&   16.2&  0.82 & 0.82&
 202.1&  201.9&  283.5 & 0.71 & 0.71\\
\hline   
\end{tabular}
\end{center}

\newpage

Table 4: $M_1$, $M_2$, $r_1$, and $r_2$ in the absence of
oscillations for collapse phase neutrino spectra obtained by
changing the parameters for the progenitor star. \\

\begin{center}
\begin{tabular}{|c|c|c|c|c|c|c|c|c|c|c|c|c|}
\hline
Progenitor & \multicolumn{6}{|c|}{$15M_\odot$}
& \multicolumn{6}{|c|}{$25M_\odot$} \\
\hline
Spectrum& \multicolumn{2}{|c|}{$M_1$} & $r_1$ &
\multicolumn{2}{|c|}{$M_2$} & $r_2$ & \multicolumn{2}{|c|}{$M_1$}
& $r_1$ & \multicolumn{2}{|c|}{$M_2$} & $r_2$ \\
\cline{2-3} \cline{5-6} \cline{8-9} \cline{11-12}
& SK & SNO & & SK & SNO & & SK & SNO & & SK & SNO & \\
\hline
Combined & 13.3 & 16.1 & .83 & 202 & 282 & .72 
      & 13.3 & 15.9 & .84 & 200 & 275 & .73 \\
\hline
Only &&&&&&&&&&&& \\
Free & 15.9 & 17.5 & .91 & 270 & 321 & .84
     & 14.9 & 16.8 & .89 & 241 & 300 & .81 \\
protons &&&&&&&&&&&& \\
\hline
Only &&&&&&&&&&&& \\
Heavy & 9.1 & 10.4 & .87 & 89 & 118 & .75 
      & 8.9 & 10.3 & .87 & 86 & 115 & .75 \\
nuclei &&&&&&&&&&&& \\
\hline
\end{tabular}
\end{center}

\newpage

Table 5: $R_{SNO}$ for collapse phase neutrinos as a function of
the mass splitting $\Delta$ for $\vartheta = 45^o$. Results are
presented for oscillation to sequential as well as sterile
neutrinos.\\

\begin{center}
\begin{tabular}{|c|c|c|}
\hline
$\Delta$ & \multicolumn{2}{|c|}{$R_{SNO}$} \\
\cline{2-3}
in & & \\
$10^{-18}$ eV$^2$ & sequential & sterile  \\
\hline
 0.0&   .431&   .431 \\
 0.3&   .844&   .441 \\
 0.6&  1.704&   .419 \\
 0.9&  1.356&   .419 \\
 1.2&   .833&   .431\\
 1.5&   .661&   .434\\
 1.8&   .708&   .434\\
 2.1&   .890&   .433\\
 2.4&  1.014&   .427 \\
 2.7&   .990&   .427 \\
 3.0&   .886&   .430 \\
 3.5&   .797&   .433 \\
 4.0&   .843&   .431 \\
 4.5&   .901&   .430 \\
 5.0&   .959&   .434 \\
 5.5&   .950&   .435 \\
 6.0&   .769&   .430 \\
 6.5&   .903&   .432 \\
\hline 
\end{tabular}
\end{center}

\newpage

Table 6: $R_{SNO}$ for collapse phase neutrinos in the absence of
oscillations for different neutrino spectra
obtained by changing the parameters of the progenitor star.\\

\begin{center}
\begin{tabular}{|c|c|c|c|}
\hline
& \multicolumn{3}{|c|} {Spectrum}\\
\cline{2-4}
Progenitor & Combined & only free protons& only heavy nuclei\\
\hline
15M$_{\odot}$ & 
0.431 & 0.445 & 0.384 \\
\hline
25M$_{\odot}$ &
0.431 & 0.440 & 0.382 \\
\hline
\end{tabular}
\end{center}

\newpage

Table 7: $M_1$ and $M_2$ for post-bounce epoch neutrinos for
different values of the mass splitting $\Delta$ for the SK and
SNO detectors.  Results are presented for oscillation to
sequential as well as sterile neutrinos.  $r_1$ and $r_2$ are
also shown for both cases. The mixing angle $\vartheta$ has been
chosen to be $45^o$.

\begin{center}
\begin{tabular}{|c|c|c|c|c|c|c|c|c|c|c|c|c|}
\hline
$\Delta$ in & \multicolumn{4}{|c|}{$M_1$}
& \multicolumn{2}{|c|}{$r_1$}
& \multicolumn{4}{|c|}{$M_2$}
& \multicolumn{2}{|c|}{$r_2$} \\
\cline{2-13}
10$^{-19}$ & \multicolumn{2}{|c|}{SK} & \multicolumn{2}{|c|}{SNO}
& seq. & st. & 
\multicolumn{2}{|c|}{SK} & \multicolumn{2}{|c|}{SNO}
& seq. & st.  \\
\cline{2-5}\cline{8-11}
eV$^2$ & seq. & st. & seq. & st.  
& & & seq. & st. & seq. & st. & & \\
\hline

  .0&     23.1&     23.1&     22.5&     22.5&      1.02&
1.02&    599.3&    599.3&    573.0&    573.0&      1.05&
1.05 \\

  .3&     24.3&     25.4&     24.3&     24.9&      1.00&
1.02&    650.1&    696.6&    649.7&    673.1&      1.00&
1.03 \\

  .6&     25.0&     26.7&     25.5&     26.2&       .98&
1.02&    688.8&    792.8&    712.8&    767.7&       .97&
1.03\\

  .9&     25.0&     20.0&     25.9&     19.1&       .97&
1.05&    696.8&    504.4&    736.2&    462.7&       .95&
1.09\\

 1.2&     25.3&     17.2&     26.4&     16.9&       .96&
1.02&    708.8&    328.2&    764.5&    315.1&       .93&
1.04\\

 1.5&     25.5&     19.4&     26.8&     19.1&       .95&
1.01&    719.8&    400.0&    786.5&    390.4&       .92&
1.02\\

 1.8&     25.4&     21.9&     26.6&     21.6&       .95&
1.02&    717.1&    518.0&    781.9&    503.5&       .92&
1.03\\

 2.1&     25.0&     23.8&     25.8&     23.4&       .97&
1.02&    696.6&    619.2&    742.1&    598.7&       .94&
1.03\\

 2.4&     24.5&     24.9&     24.7&     24.3&       .99&
1.02&    666.8&    685.1&    682.4&    658.3&       .98&
1.04\\

 2.7&     24.1&     25.1&     24.0&     24.4&      1.00&
1.03&    643.6&    708.1&    637.9&    676.0&      1.01&
1.05\\

 3.0&     24.0&     24.6&     24.0&     23.9&      1.00&
1.03&    638.8&    688.9&    633.0&    654.4&      1.01&
1.05\\

 3.5&     24.6&     22.8&     25.1&     22.2&       .98&
1.03&    666.8&    596.8&    690.7&    567.2&       .97&
1.05\\

 4.0&     25.2&     21.6&     26.2&     21.2&       .96&
1.02&    702.7&    521.0&    754.2&    500.5&       .93&
1.04\\

 4.5&     25.4&     22.0&     26.5&     21.5&       .96&
1.02&    714.5&    531.2&    773.3&    512.0&       .92&
1.04\\

 5.0&     25.0&     23.2&     25.9&     22.6&       .97&
1.02&    697.4&    594.9&    742.4&    570.7&       .94&
1.04\\

 5.5&     24.6&     24.0&     25.0&     23.4&       .98&
1.03&    669.6&    647.8&    691.3&    619.2&       .97&
1.05\\

 6.0&     24.3&     23.8&     24.6&     23.1&       .99&
1.03&    653.3&    642.9&    664.3&    612.4&       .98&
1.05\\

 6.5&     24.4&     22.8&     24.9&     22.2&       .98&
1.03&    662.4&    596.6&    684.3&    567.5&       .97&
1.05\\

 7.0&     25.1&     23.0&     26.0&     22.6&       .97&
1.02&    696.1&    587.8&    739.8&    564.6&       .94&
1.04\\

 7.5&     25.2&     22.3&     26.1&     21.8&       .96&
1.02&    703.2&    555.3&    753.6&    532.8&       .93&
1.04\\
\hline
\end{tabular}
\end{center}

\newpage

Table 8: Range of variation of $M_1$, $M_2$, $M_3$, $r_1$, $r_2$
and $r_3$  for post-bounce epoch neutrinos in the absence of
oscillations obtained by changing the input spectra $f^e(E)$,
$\bar{f}^e(E)$, and $f^x(E)$.  The symbol + (--) indicates an
increase (decrease) of the corresponding spectrum by 30\% while 0
signifies the unchanged spectrum.\\

\begin{center}
\begin{tabular}{|c|c|c|c|c|c|c|c|c|}
\hline
$f^e(E)$& ${\bar f}^e (E)$&  $f^x(E) $ &   
\multicolumn{2}{|c|} {$M_1$} & $r_1$ & \multicolumn{2}{|c|}
{$M_2$} & $r_2$  \\ \cline{4-5} \cline{7-8}
 & & & SK & SNO & & SK & SNO & \\
\hline 
0 & 0 & 0 &  
23.09 &  22.53 & 1.02 &  599.26 & 572.97 &  1.05 \\
+ & + & -- &   
    23.09 & 22.53 & 1.02 & 599.26 & 572.97 &  1.05\\
+ & -- & -- &  
    23.05 & 21.70 & 1.06 &   597.57 & 534.95 &  1.12\\
+ & -- & + &  
    23.05 & 21.70 & 1.06 &   597.57 & 534.95 &  1.12\\
-- & + & + &  
    23.12 & 23.37 & 0.99 &   600.18 & 610.98 &  0.98\\
-- & + & -- &  
    23.12 & 23.37 & 0.99 &   600.18 & 610.98 &  0.98\\
-- & -- & + &  
    23.09 & 22.53 & 1.02 &   599.26 & 572.97 &  1.05\\
\hline
\end{tabular}
\end{center}

\newpage

Table 9: $R_{SNO}$ for post-bounce epoch neutrinos as a function
of the mass splitting $\Delta$ for $\vartheta = 15^o$ and $45^o$.
Results are presented for oscillation to sequential as well as
sterile neutrinos.\\

\begin{center}
\begin{tabular}{|c|c|c|c|c|}
\hline
$\Delta$ & \multicolumn{4}{|c|}{$R_{SNO}$} \\
\cline{2-5}
in & \multicolumn{2}{|c|}{$\vartheta = 15^0$} &
\multicolumn{2}{|c|}{$\vartheta = 45^0$} \\
\cline{2-5}
$10^{-19}$ eV$^2$ & sequential & sterile & sequential & sterile \\
\hline
  .0&   1.929&   1.929&   1.929&   1.929 \\ 
  .3&   1.952&   2.050&   2.024&   2.584 \\
  .6&   1.894&   2.212&   1.796&   4.837\\
  .9&   1.796&   2.247&   1.488&   6.007\\
 1.2&   1.749&   2.201&   1.368&   4.565\\
 1.5&   1.765&   2.145&   1.406&   3.552\\
 1.8&   1.817&   2.108&   1.549&   3.101\\
 2.1&   1.876&   2.093&   1.733&   2.945\\
 2.4&   1.915&   2.094&   1.876&   2.951\\
 2.7&   1.923&   2.104&   1.908&   3.055\\
 3.0&   1.905&   2.118&   1.836&   3.211\\
 3.5&   1.850&   2.139&   1.647&   3.469\\
 4.0&   1.814&   2.145&   1.539&   3.555\\
 4.5&   1.818&   2.137&   1.552&   3.444\\
 5.0&   1.851&   2.125&   1.652&   3.292\\
 5.5&   1.882&   2.119&   1.753&   3.219\\
 6.0&   1.881&   2.116&   1.751&   3.186\\
 6.5&   1.856&   2.118&   1.667&   3.210\\
 7.0&   1.844&   2.143&   1.628&   3.523\\
 7.5&   1.830&   2.136&   1.585&   3.430\\
\hline
\end{tabular}
\end{center}

\newpage

Table 10: Range of variation of $R_{SNO}$ for post-bounce epoch
neutrinos in the absence of oscillations obtained by changing the
input spectra $f^e(E)$, $\bar{f}^e(E)$, and $f^x(E)$. The symbol
+ (--) indicates an increase (decrease) of the corresponding
spectrum by 30\% for case (a) and 10\% for case (b) while 0
signifies the unchanged spectrum.\\

\begin{center}
\begin{tabular}{|c|c|c|c|c|}
\hline
& & & \multicolumn{2}{|c|}{$R_{SNO}$} \\
\cline{4-5}
$f^e(E)$& ${\bar f}^e (E)$&  $f^x(E) $ &   
Case (a) & Case (b) \\
&&& $\pm 30\%$ & $\pm 10\%$ \\
\hline
0 & 0 & 0 &  
1.93 & 1.93 \\
+ & + & -- &   
1.29 & 1.68 \\
+ & -- & -- &  
1.48 & 1.78 \\
+ & -- & + &  
2.32 & 2.06 \\
-- & + & + &  
2.37 & 2.08 \\
-- & + & -- &  
1.54 & 1.80 \\
-- & -- & + &  
3.12 & 2.24 \\
\hline
\end{tabular}
\end{center}

\newpage

\begin{center}
{\bf{\Large Figure Captions}}
\end{center}
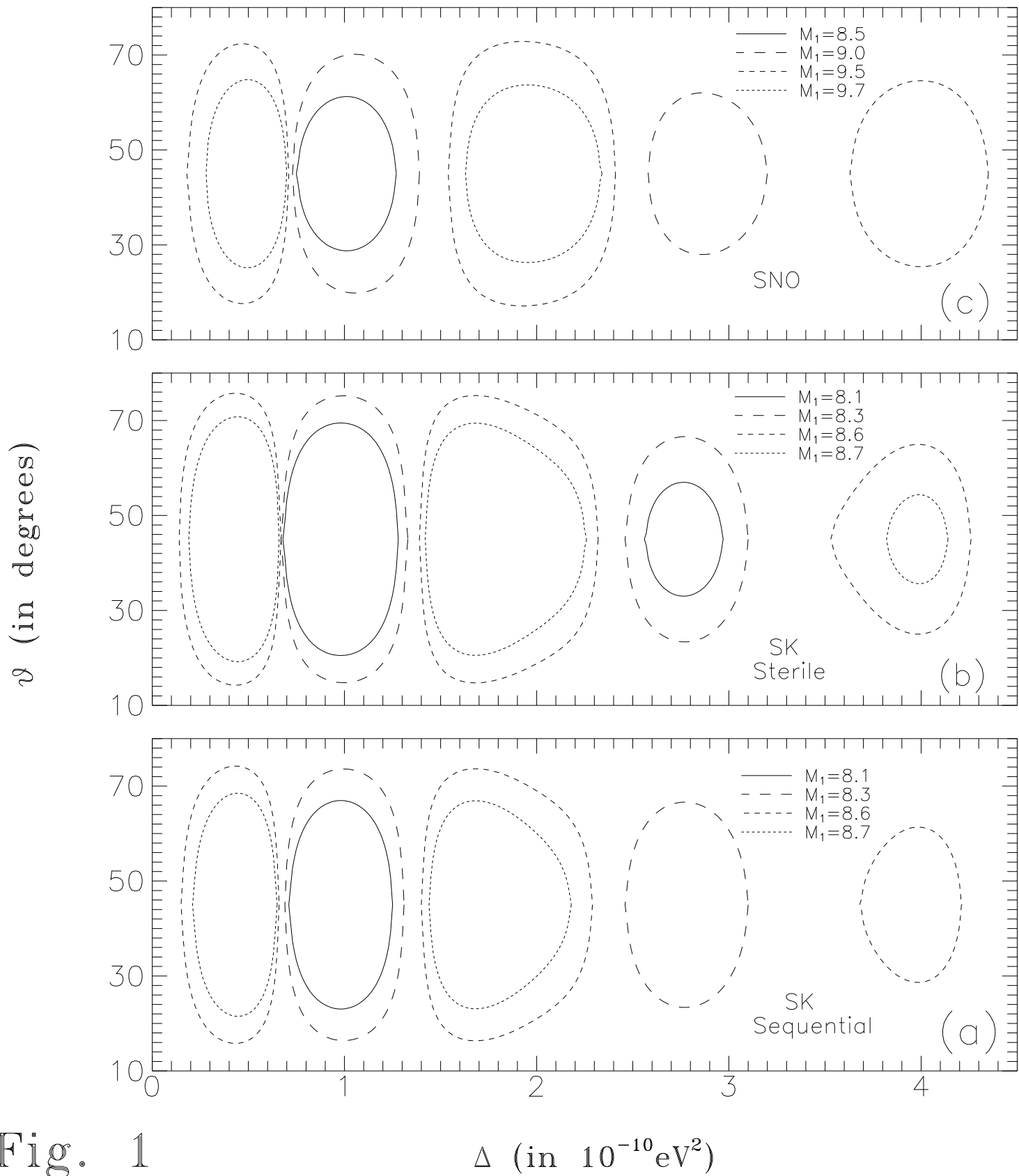
Fig. 1: Contours of constant $M_1$ for solar neutrinos in the
$\Delta$--$\vartheta$ plane. (a) and (b) correspond to
oscillation to sequential and sterile neutrinos respectively at
SK while (c) corresponds to SNO where the two scenarios give
identical results.\\
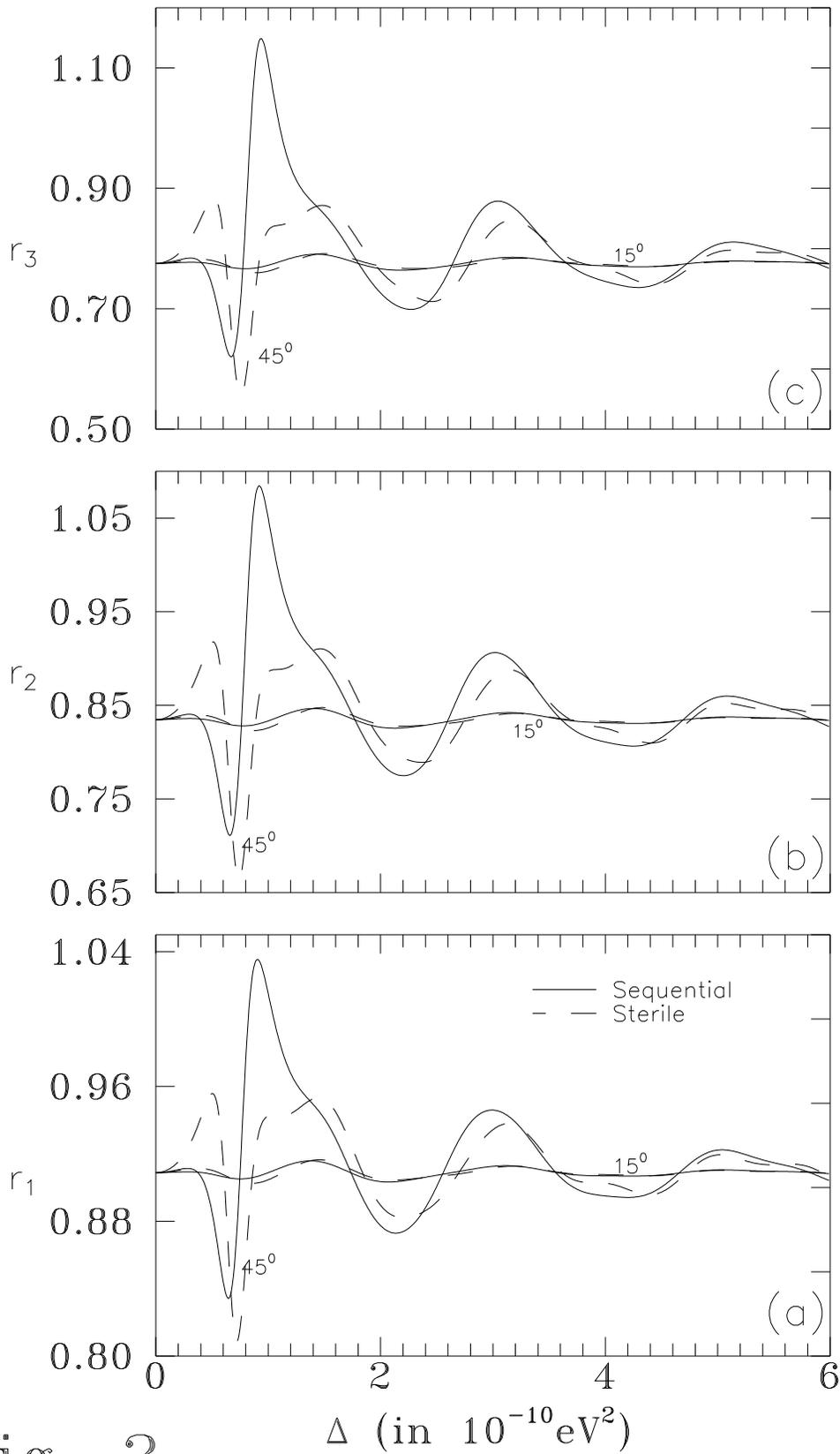
Fig. 2: The variables (a) $r_1$, (b) $r_2$, and (c) $r_3$ for
solar neutrinos as a function of the mass splitting $\Delta$ for
two values of the mixing angle $\vartheta$ = $45^o$ and $15^o$.
The solid (broken) curves correspond to oscillation to sequential
(sterile) neutrinos.\\
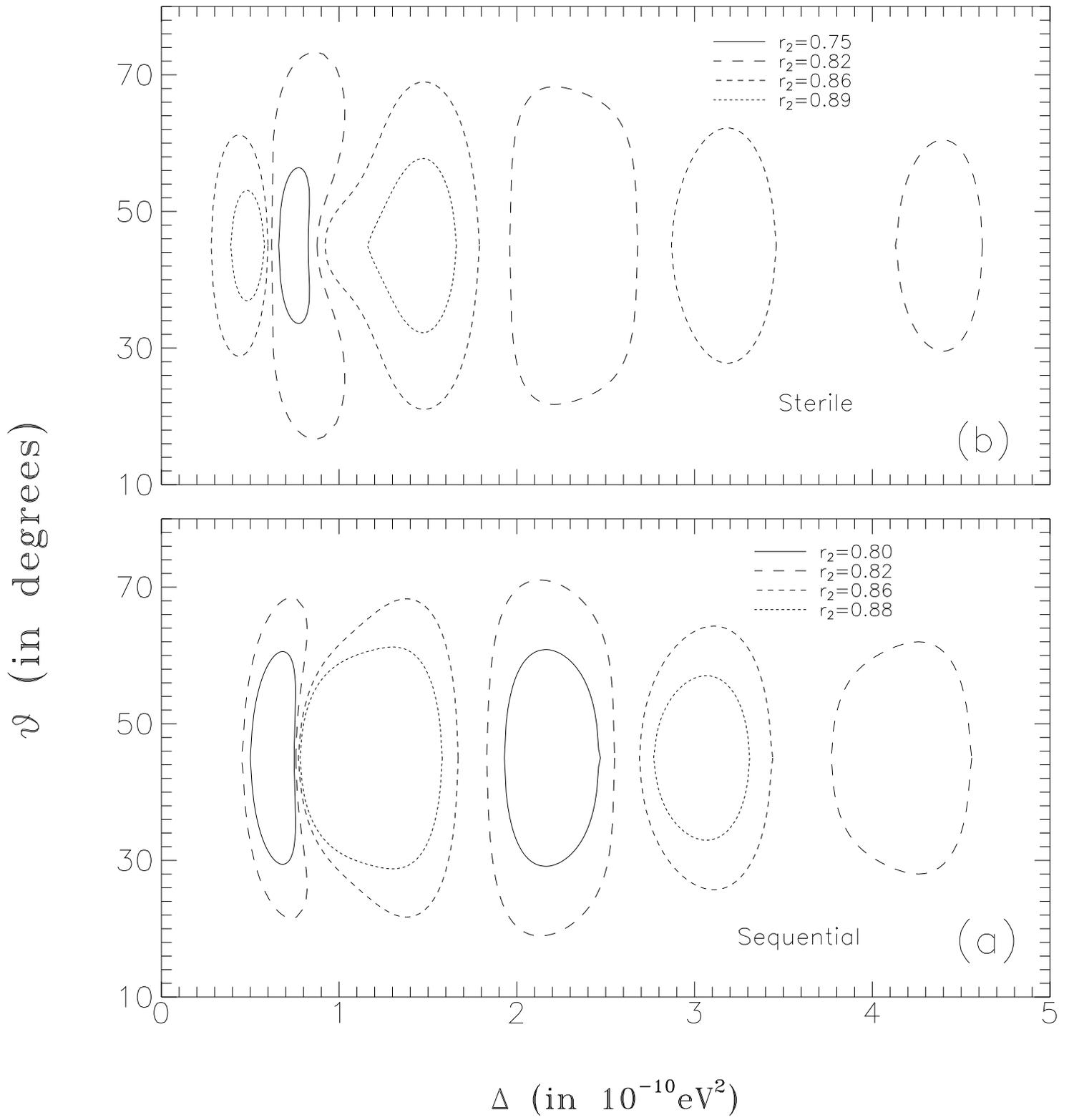
Fig. 3: Contours of constant $r_2$ for solar neutrinos in the
$\Delta$--$\vartheta$ plane. (a) and (b) correspond to
oscillation to sequential and sterile neutrinos respectively. In
the absence of oscillation  $r_2$ = 0.83.\\
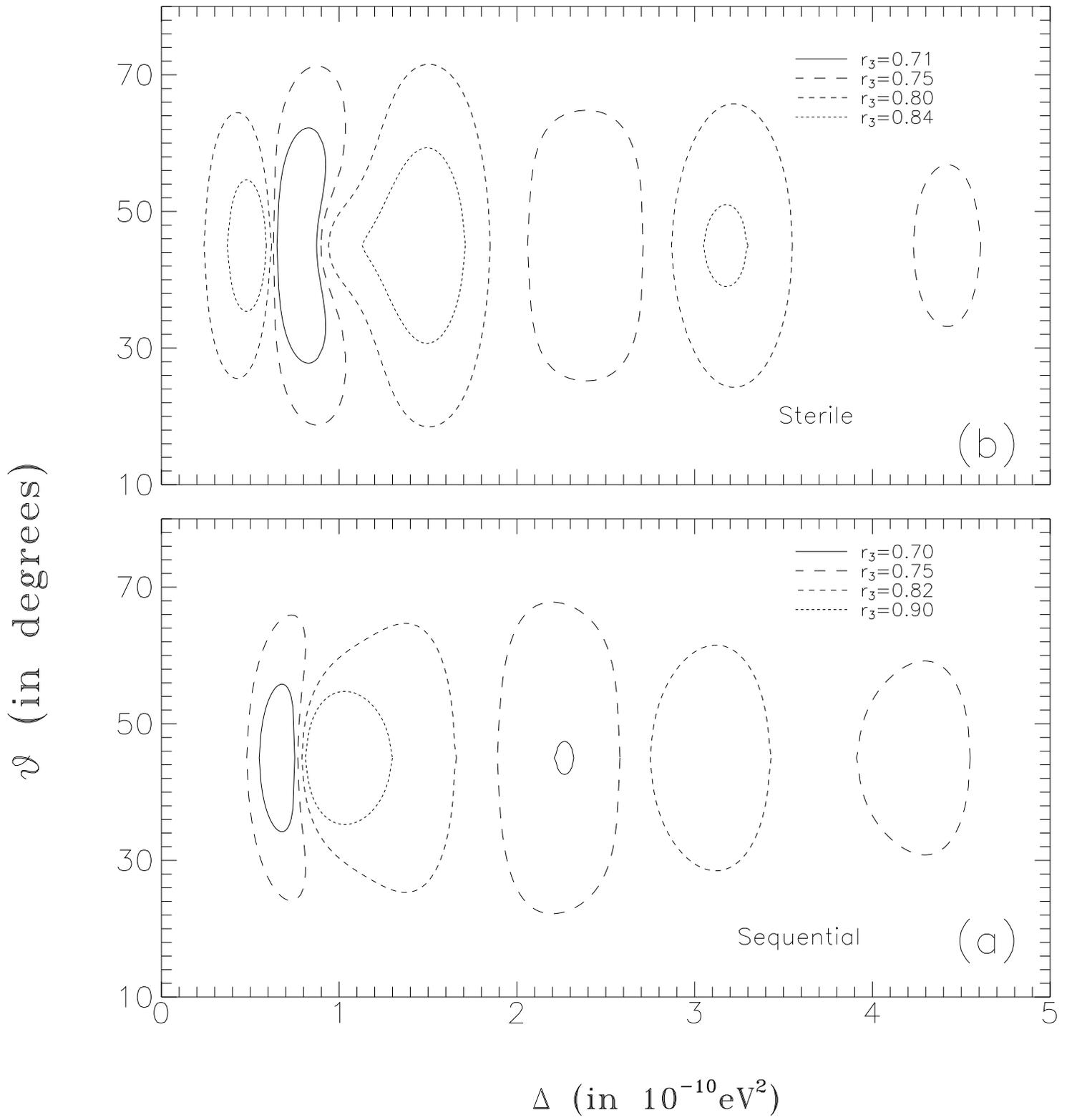
Fig. 4: Contours of constant $r_3$ for solar neutrinos in the
$\Delta$--$\vartheta$ plane. (a) and (b) correspond to
oscillation to sequential and sterile neutrinos respectively. In
the absence of oscillation $r_3$ = 0.78.\\
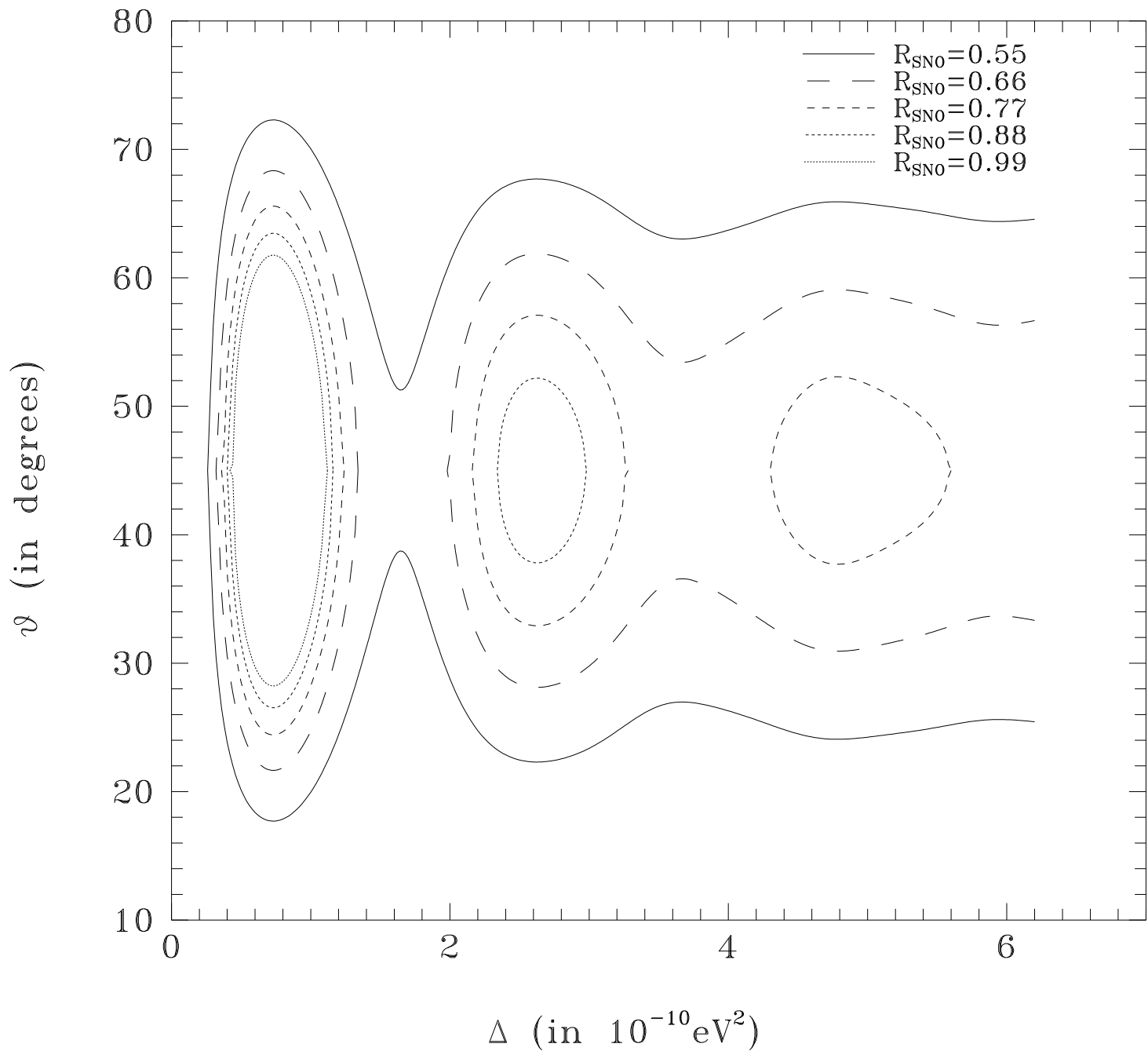
Fig. 5: Contours of constant $R_{SNO}$ in the
$\Delta - \vartheta$ plane for oscillation of solar neutrinos to
sequential neutrinos.  In the absence of  oscillation $R_{SNO}$ =
0.38.  \\
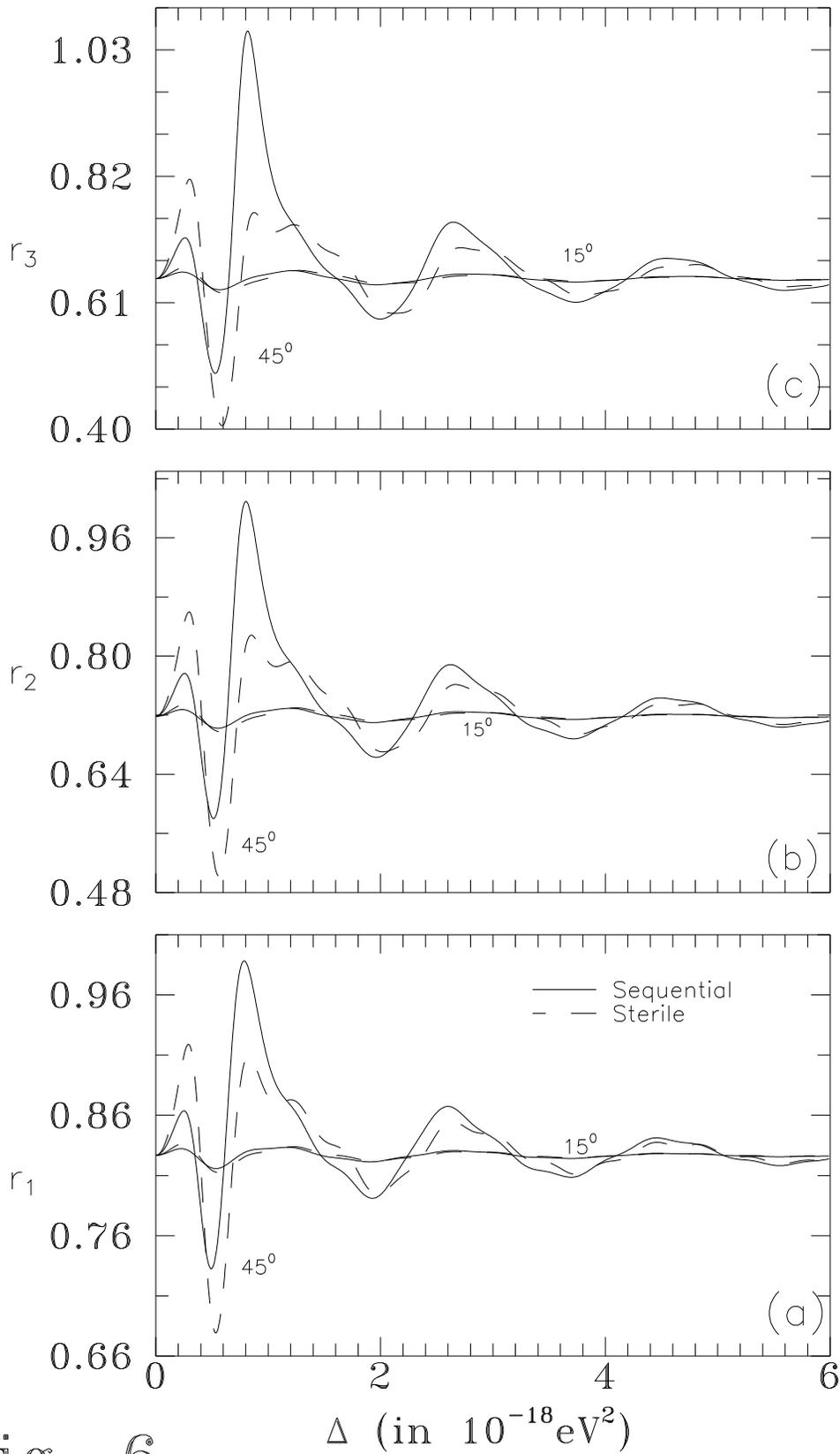
Fig. 6: The variables (a) $r_1$, (b) $r_2$, and (c) $r_3$ for
collapse phase neutrinos as a function of the mass splitting
$\Delta$ for two values of the mixing angle $\vartheta$ = $45^o$
and $15^o$. The solid (broken) curves correspond to oscillation to
sequential (sterile) neutrinos.\\
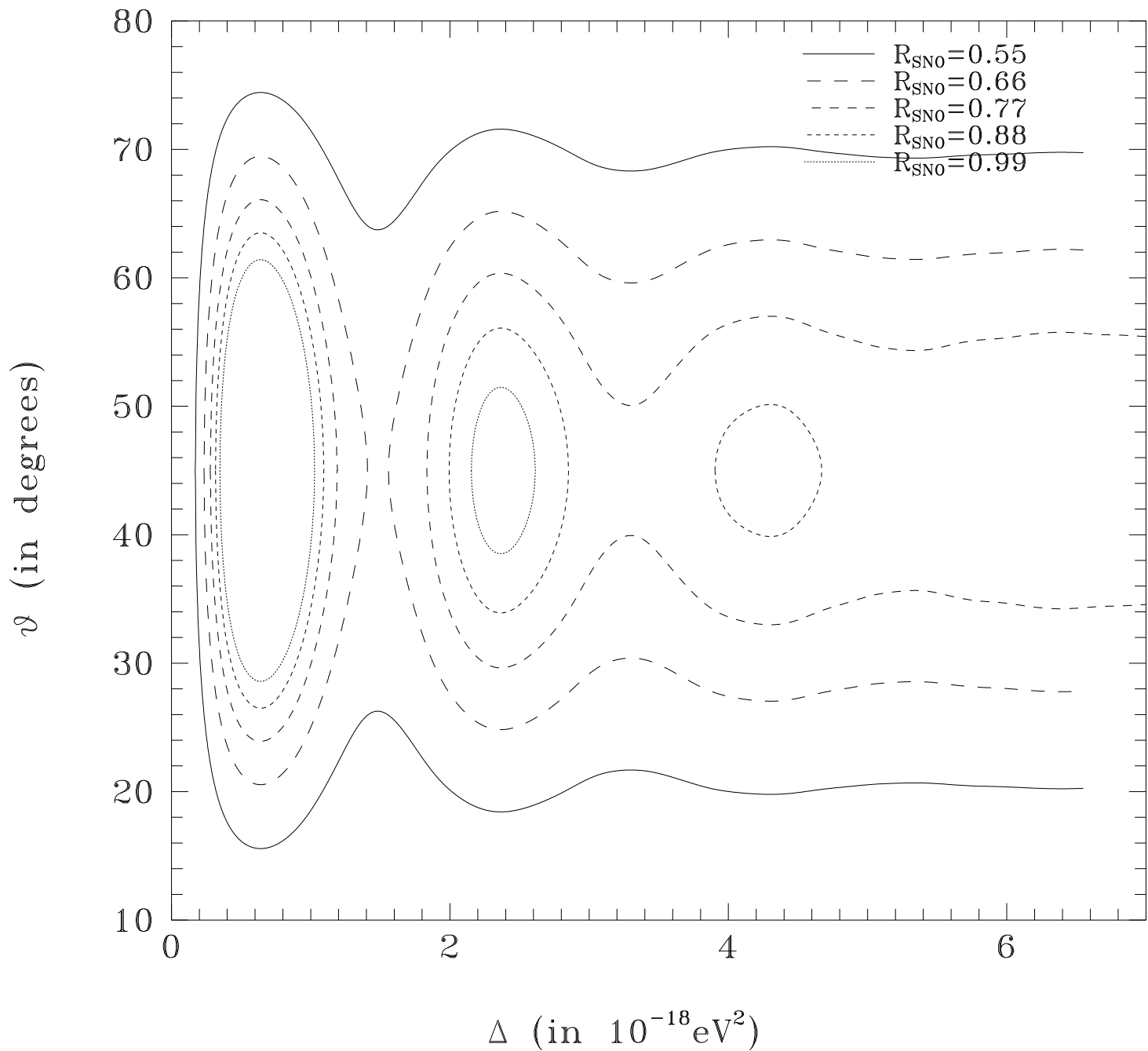
Fig. 7: Contours of constant $R_{SNO}$ in the $\Delta -
\vartheta$ plane for oscillation of collapse phase 
neutrinos to sequential neutrinos. In the absence of oscillation
$R_{SNO}$ = 0.43.  \\ 
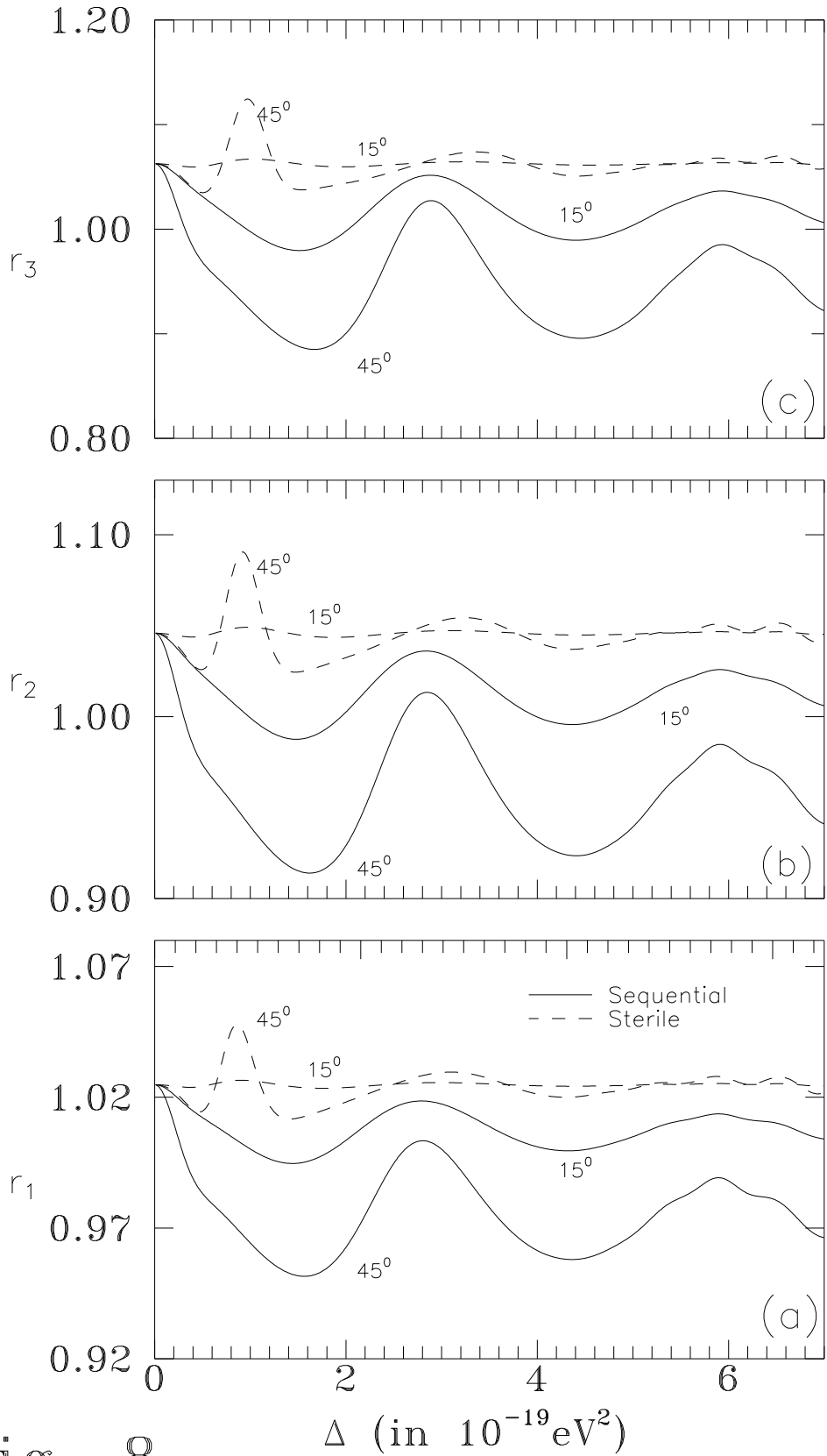
Fig. 8: The variables (a) $r_1$, (b) $r_2$, and (c) $r_3$ for
post-bounce epoch neutrinos as a function of the mass splitting
$\Delta$ for two values of the mixing angle $\vartheta$ = $45^o$
and $15^o$. The solid (broken) curves correspond to oscillation
to sequential (sterile) neutrinos.\\
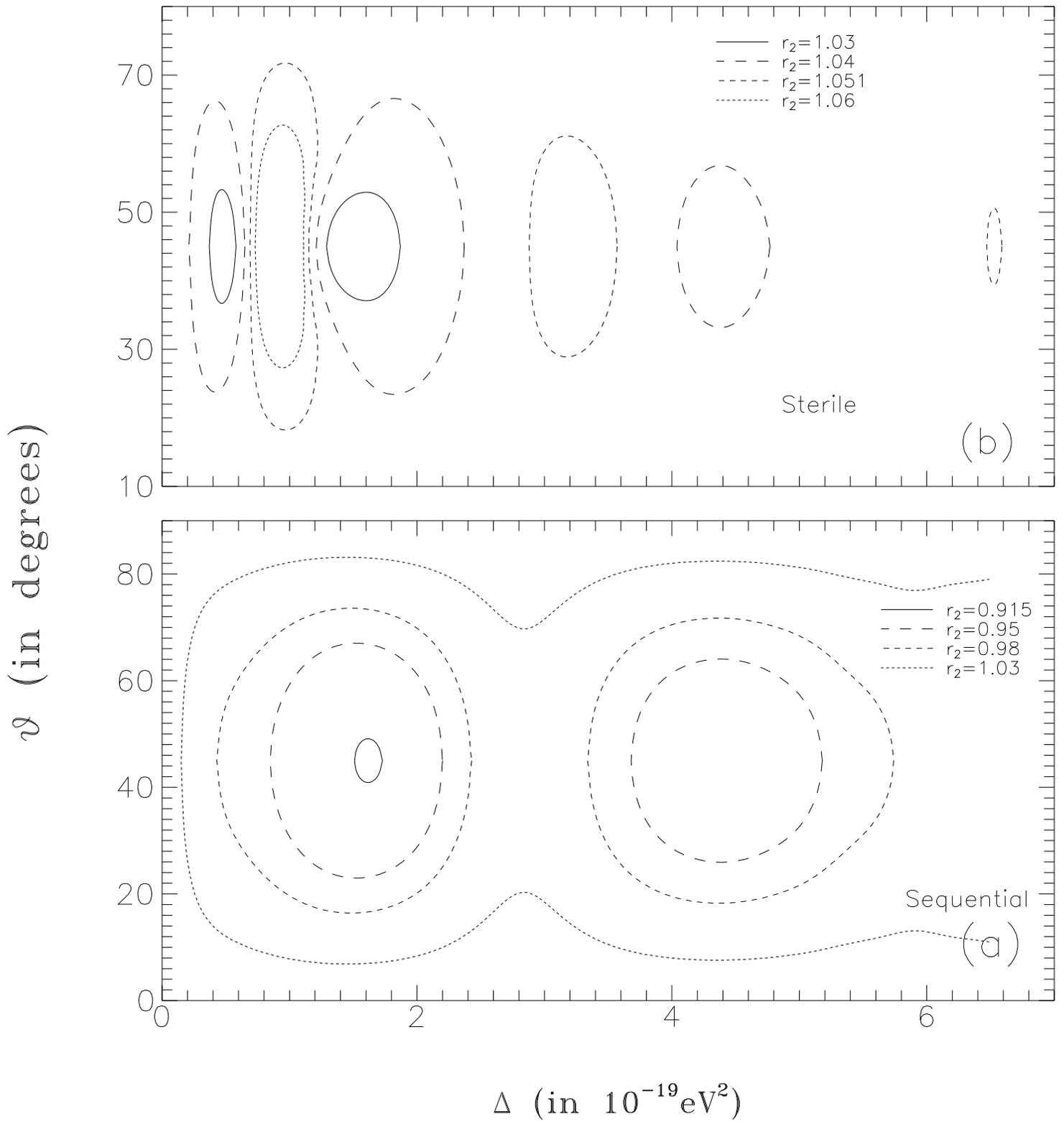
Fig. 9: Contours of constant $r_2$ for post-bounce epoch
neutrinos in the $\Delta$--$\vartheta$ plane. (a) and (b)
correspond to oscillation to sequential and sterile neutrinos
respectively. In the absence of oscillation $r_2$ = 1.05.\\
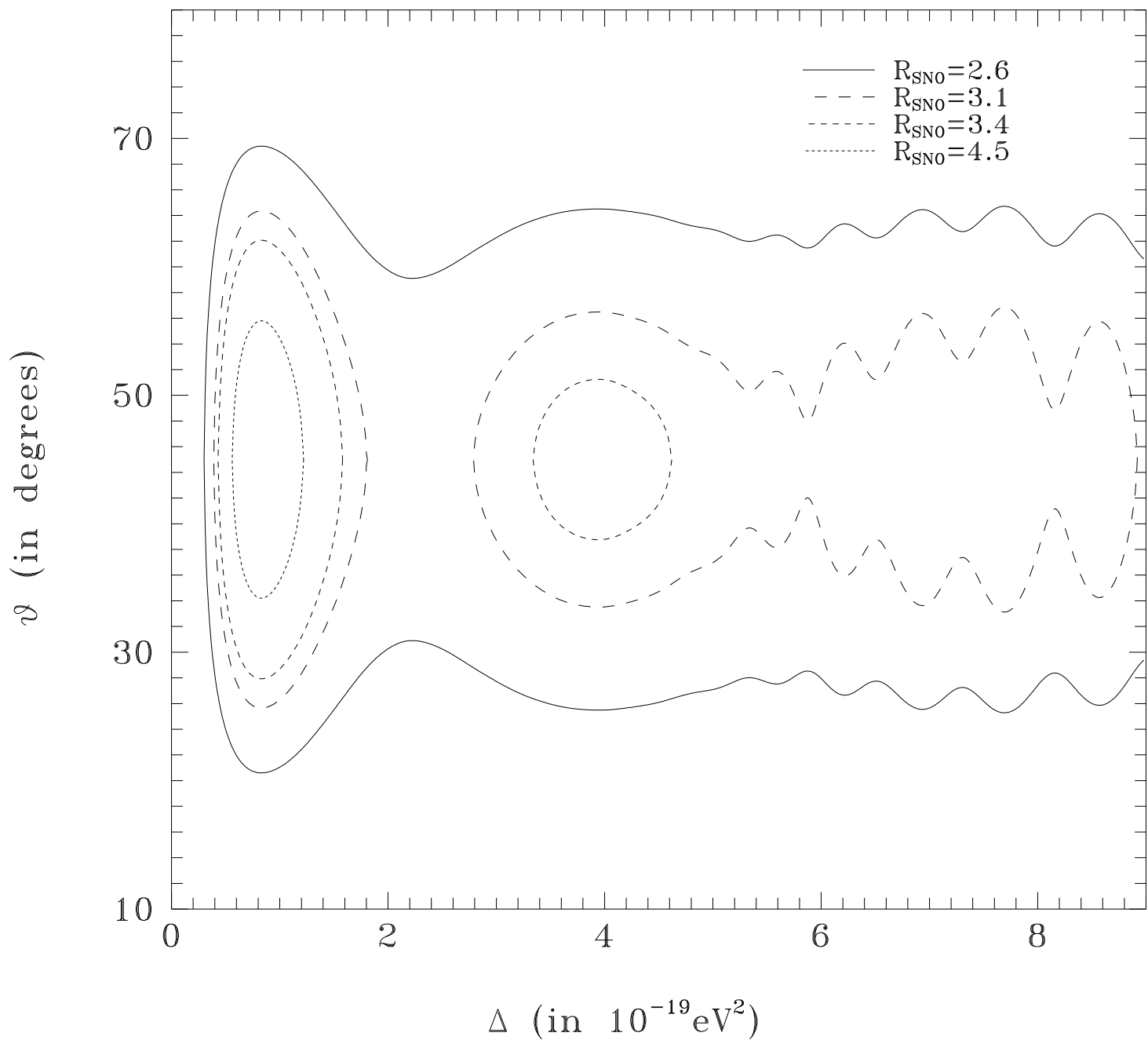
Fig. 10: Contours of constant $R_{SNO}$ in the $\Delta -
\vartheta$ plane for oscillation of post-bounce epoch 
neutrinos to sterile neutrinos. In the absence of oscillation
$R_{SNO}$ = 1.93.

\end{document}